\documentclass[11pt, dvipsnames]{article}

\usepackage[margin=1.0in]{geometry}

\RequirePackage{amsmath}
\RequirePackage{natbib}
\usepackage[colorlinks,citecolor=blue,urlcolor=blue, linkcolor = blue ]{hyperref}
\usepackage{cleveref}
\usepackage{tikz,pgf}
\usepackage{multirow, multicol}
\usepackage{float}
\usepackage{amsfonts}
\usepackage{lscape}
\usepackage{pdflscape}
\usepackage[normalem]{ulem}
\usepackage{bm}
\usepackage[font=singlespacing]{caption}
\newcommand{\yl}[1]{\textcolor{black}{#1}}

\def\independent{\perp\!\!\!\perp}

\def\spacingset#1{\renewcommand{\baselinestretch}%
{#1}\small\normalsize} \spacingset{1}

\usepackage{ulem}

\newcommand{\blind}{1}

\begin{document}

\def\spacingset#1{\renewcommand{\baselinestretch}%
{#1}\small\normalsize} \spacingset{1}

%%%%%%%%%%%%%%%%%%%%%%%%%%%%%%%%%%%%%%%%%%%%%%%%%%%%%%%%%%%%%%%%%%%%%%%%%%%%%%

\if1\blind
{
  \title{ Partially Pooled Propensity Score Models for Average Treatment Effect Estimation with Multilevel Data}
  \author{Youjin Lee\thanks{This work was conducted while Dr. Lee was a post-doctoral fellow at Johns Hopkins Bloomberg School of Public Health. Email: youjin.lee@pennmedicine.upenn.edu},  ~$\text{Trang Q. Nguyen}^\dagger$, and $\text{Elizabeth A. Stuart}^\dagger$ \\ 
  $\text{University of Pennsylvania}^{*}$ and $\text{Johns Hopkins Bloomberg School of Public Health}^\dagger$}
    %Center for Causal Inference, University of Pennsylvania,\\
    %Edward Kennedy \\
    %Department of Statistics, Carnegie Mellon University,\\
    %and \\
    %Nandita Mitra\\
    %Department of Biostatistics, Epidemiology \& Informatics, University of Pennsylvania}
  \maketitle
} \fi

\if0\blind
{
  \bigskip
  \bigskip
  \bigskip
  \begin{center}
    {\LARGE\bf Title}
\end{center}
  \medskip
} \fi

\begin{abstract}
Causal inference analyses often use existing observational data, which in many cases has some clustering of individuals. In this paper we discuss propensity score weighting methods in a multilevel setting where within clusters individuals share unmeasured confounders that are related to treatment assignment and the potential outcomes. We focus in particular on settings where \yl{models with fixed cluster effects} are either not feasible or not useful due to the presence of a large number of small clusters.
We found, both through numerical experiments and theoretical derivations, that a strategy of grouping clusters with similar treatment prevalence and estimating propensity scores within such cluster groups is effective in reducing bias from unmeasured cluster-level covariates under mild conditions on the outcome model. We apply our proposed method in evaluating the effectiveness of center-based pre-school program participation on children's achievement at kindergarten, using the Early Childhood Longitudinal Study, Kindergarten data. 
\end{abstract}

\vfill
%\spacingset{1.45} % DON'T change the spacing!

%\begin{keyword}
%\kwd{heterogeneous treatment effect} \kwd{propensity scores} 
%\kwd{weighting} 
%\kwd{unmeasured confounder}
%\end{keyword}

%\end{frontmatter}
%\spacingset{1.45} % DON'T change the spacing!

\section{Introduction}
\label{sec:intro}

Human subjects are often clustered in communities, schools, hospitals, online social groups, etc., sharing the same environmental factors, services, interventions, or physical facilities. This clustering also often makes data collection more convenient as subjects are close to one another in physical or virtual space~\citep{rabe2006multilevel, arpino2011specification, leite2015evaluation}. As a result, clustered observational data are widely used to investigate causal relationships between treatments and outcomes that occur at the individual level, but often without complete knowledge of shared characteristics and contextual backgrounds.

When a cluster-level characteristic influences both the treatment and outcome of interest, it \textit{confounds} the causal effect~\citep{greenland1999confounding} by inducing a spurious association between the two variables. In addition, a cluster-level characteristic may also bring about systematic variation across clusters in the direction and/or magnitude of the causal effect~\citep{have2004deviations, vanderweele2007four}, in which case it is called an \textit{effect modifier}. 
\yl{In this paper, we consider the effect of center-based pre-school education on math proficiency in elementary school as a motivating example. We estimate the average causal effect based on the sample of students clustered in multiple schools. In this case, school-level covariates (e.g., the region's educational infrastructure) may confound the relationships between the treatment and the outcome and also modify the treatment effect. For instance, the center-based pre-school education effect might be more evident for schools in particular regions than those in other areas due to differences in attitude about education or in the resources put into schools in that region.}

Estimating an average treatment effect via propensity score analysis requires balancing the distribution of confounders and effect modifiers between different treatment arms to remove bias in the causal estimate.
Unfortunately, cluster-level characteristics are often unmeasured and thus not balanced between different arms~\citep{thoemmes2011use,  yang2018propensity, he2018inverse}, which could lead to bias. This is called the \textit{unmeasured context} problem~\citep{arpino2011specification}. Here we focus on how to tackle bias due to unmeasured contextual factors in propensity score analysis with clustered observational data.

Arguably, the ideal way to eliminate bias due to unmeasured cluster-level characteristics is to separately fit one propensity score model for each cluster and to do propensity score matching, weighting, or subclassification separately within each cluster~\citep{kim2015multilevel} -- we call this \textit{cluster-specific propensity score estimation and use}. 
However, in many multilevel datasets, the cluster sample sizes are too small or too variable to allow doing this for each cluster. 
For instance, in the Early Childhood Longitudinal Study's Kindergarten (ECLS-K) dataset (a longitudinal nationally representative cohort of children followed starting from kindergarten~\citep{tourangeau2009early}), the first, second, and third quartiles of cluster (school) sample size are 1, 18, and 22. Over a quarter of the children in the sample form their own singleton clusters. About 4\% of the participating schools have only 2-5 children in the sample, and another 25\% of the participating schools have 6-20 children in the sample. These cluster sample sizes are not sufficient for the number of covariates that we would usually need to adjust for. It is thus no surprise that common practice is to estimate a single propensity score model pooling observations from all the clusters and use them over all individuals in the data (which we call \textit{fully pooled propensity score estimation and use}), even though this often suffers from confounding by unmeasured context.

\yl{Small cluster sizes can lead the perhaps default fixed effect approach to not be feasible.}
In this paper, we propose an alternative strategy that overcomes this small cluster sample size problem by grouping clusters with similar treatment prevalence into several \textit{cluster groups} and then estimating propensity scores separately within each group. We call this \textit{partially pooled propensity score estimation}. \yl{Once the clusters are grouped, any propensity score estimation appraoch can be used within those groups}. This strategy allows variation in propensity score models across cluster groups, resulting in better control for unmeasured contextual factors and less biased causal effect estimates \yl{under mild assumptions on the outcome model}.

The paper proceeds as follows. Section~\ref{sec:settingetc} describes the setting, introduces the notation, and presents the estimand and assumptions. Section~\ref{sec:motivation} introduces the data example and presents an exploratory analysis of a cluster-level covariate to motivate the proposed method. Section~\ref{sec:review} reviews related work on propensity score methods with clustered data and unmeasured cluster-level confounding. Section~\ref{sec:method} presents our proposed propensity score estimation method and the theoretical properties of two relevant inverse probability weighting estimators. Section~\ref{sec:simulation} reports simulation results on the proposed method.
Our method is finally applied to the ECLS-K data in evaluating the effect of center-based pre-school education on kindergarten math scores in Section~\ref{sec:application}. We conclude the paper with a discussion. All the relevant code can be found at the first author's github repository\footnote{\url{https://github.com/youjin1207/PartialIPW}}.

\section{Setting, estimand, and assumptions}
\label{sec:settingetc}

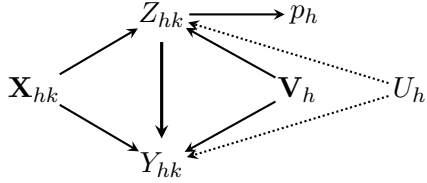
\begin{figure}
\begin{centering}
	\begin{tikzpicture}[>=stealth, node distance=0.5cm]
	\tikzstyle{format} = [thin, circle, minimum size=5.0mm,
	inner sep=0.1pt]
	\tikzstyle{square} = [thin, rectangle, draw]
	\begin{scope}[xshift=0.0cm, yshift = 0cm]
	\path[->, thick]
	node[format, yshift=1cm,  circle, line width=0.3mm] (x) {$\mathbf{X}_{hk}$}
	node[format, right of=x, xshift = 1.2cm, yshift = 1cm,  circle, line width=0.3mm] (z) {$Z_{hk}$}
	node[format, right of=z, xshift = 1.4cm, yshift = 0cm,  circle, line width=0.3mm] (p) {$p_{h}$}
	node[format, right of=x, xshift = 1.2cm, yshift = -1cm, circle, line width=0.3mm] (y) {$Y_{hk}$}
	node[format, right of=x, xshift = 3cm, circle, line width=0.3mm] (v) {$\mathbf{V}_{h}$}
	node[format, right of=v, xshift = 1cm, circle, line width=0.3mm] (u) {$U_{h}$}
	
	%node[format, below of = y, yshift = -0.1cm] (c) {(a) Missing cluster-level confounders}
	
	(x) edge[thick, black] (z)
	(x) edge[thick, black] (y)
	(z) edge[very thick] (y)
	(z) edge[black] (p)
	(v) edge[thick, black] (z)
	(v) edge[thick, black] (y)
	(u) edge[thick, densely dotted] (y)
	(u) edge[thick, densely dotted] (z)
	;
	\end{scope}
	\end{tikzpicture} 
	\par\end{centering}
\caption{\label{fig:diagram} Directed acyclic graph showing the relationships between treatment $Z_{hk}$, outcome $Y_{hk}$, observed covariates $\mathbf{X}_{hk}$ and $\mathbf{V}_{h}$, and unobserved \yl{cluster-level} covariates $U_{h}$. The unobserved $U_h$ may \yl{be associated with} the distribution of $Z_{hk}$ and $Y_{hk}$ as well as the effect of $Z_{hk}$ on $Y_{hk}$, but these are unobserved ($\dashrightarrow$) in $\mathbf{O}_{n}$. \yl{$p_{h}$ denotes the treatment prevalence of cluster $h$, which is associated with the observed covariates and outcome through $Z_{hk}$.}}
\end{figure}

We consider clustered data in a two-level structure comprised of multiple clusters, each with possibly multiple individuals.
Given this two-level structure, we index observations using dual subscripts: $h$ indexing the cluster ($h=1,2,\dots,H$, where $H$ is the number of clusters in the population) and $k$ indexing the individual within cluster ($k=1,2,\dots,n_h$, where $n_h$ is the sample size from cluster $h$). For individual $k$ in cluster $h$, $Z_{hk}$ denotes the treatment, and $Y_{hk}$ the outcome. Here treatment and outcome are both at the individual level. Let us assume that treatment is binary, i.e., $Z_{hk}=1$ if treated, and $=0$ if not.
$\mathbf{X}_{hk} \in \mathbb{R}^{p}$ and $\mathbf{V}_h \in \mathbb{R}^{q}$ are observed pre-treatment covariates at the individual-level and cluster-level, respectively. Both $\mathbf{X}_{hk}$ and $\mathbf{V}_{h}$ can act as confounders, influencing both $Z_{hk}$ and $Y_{hk}$ as depicted in the causal diagram in Figure~\ref{fig:diagram}. Let $C_{hk}$ denote the cluster the individual belongs to; that is, $C_{hk}=h$. In summary, we are given a set of observations $\mathbf{O}_{n} = \{ ( Y_{hk}, Z_{hk}, \mathbf{X}_{hk}, \mathbf{V}_{h}, C_{hk} ) : h = 1,2,\ldots, H;~k=1,2,\ldots, n_{h},~n = \sum_{h=1}^{H} n_{h} \}$.

In addition to these observed variables, $U_h$ denotes an unobserved cluster-level variable that is a confounder and possibly also an effect modifier. $U_{h}$ may represent the cluster's contextual (e.g., a school's distance from the nearest city) or compositional (e.g., percent of the school's children who live under the poverty line) characteristics that are not captured in the dataset.

To formally represent our target estimand, we use the potential outcomes framework~\citep{rubin1974estimating, holland1986statistics}. We assume the Stable Unit Treatment Value Assumption (SUTVA) \citep{Rubin1980}, which includes no interference (i.e., the treatment of one individual does not affect the outcome of others) and irrelevance of treatment variation. 
In multilevel setting, the SUTVA  needs a careful attention; in our case, we assume that outcomes observed within cluster are possibly correlated with each other \yl{due to shared cluster level confounders}, but the treatment received by one subject does not have a direct effect on the outcome of others even in the same cluster; thus, SUTVA holds. 
Under this assumption, each individual has a potential outcome $Y_{hk}(1)$ if treated and a potential outcome $Y_{hk}(0)$ if untreated. 
The treatment effect for the individual is defined as the difference between these two potential outcomes, $\tau_{hk} := Y_{hk}(1)-Y_{hk}(0)$.
%\begin{equation}
%    \label{eq:individualeffect}
%    \tau_{hk}:=Y_{hk}(1)-Y_{hk}(0),
%\end{equation}
%which is unidentified as we do not observe both potential outcomes for an individual. 
Our target estimand, $\tau$, is the \textit{average treatment effect} (ATE) for the population \yl{\textit{conditional on current cluster membership}}, i.e., the average of the individual treatment effects with equal weights over all the individuals in the population \yl{conditional on the observed cluster assignments}~\citep{hong2006evaluating}:
%\begin{equation}
%\begin{split}
%\label{eq:ate}
%   \tau  &:= \mathbb{E}[\tau_{hk}],
%\end{split}
%\end{equation}
\begin{equation}
\label{eq:atepractical}
  \tau  = \mathbb{E}\left[Y_{hk}(1)-Y_{hk}(0)\right].
\end{equation}

\smallskip

Here the target estimand $\tau$ would be identified if we were to observe $Z_{hk},Y_{hk},\mathbf{X}_{hk},\mathbf{V}_{hk}$, and also $U_h$. This is based on the following assumptions:
\begin{enumerate}
    \item \textit{Consistency} \citep{cole2009consistency}:  
    %(which is also the treatment variation irrelevance component of SUTVA), formally
    $Y_{hk}=Y_{hk}(z)$ if $Z_{hk}=z$ for $z=0,1$;
    \item \textit{Positivity} \citep{hernan2006estimating}: 
    %formally 
    $0<p(Z_{hk}=1\mid\mathbf{X}_{hk}=\mathbf{x},\mathbf{V}_h=\mathbf{v},U_h=u)<1$ for all $(\mathbf{x},\mathbf{v},u)$ in the support of $(\mathbf{X}_{hk},\mathbf{V}_h,U_h)$;
    \item \textit{Unconfoundeness} \citep{Imbens2008}: %formally 
    $Y_{hk}(z)\independent Z_{hk}\mid \mathbf{X}_{hk},\mathbf{V}_h,U_h$ for $z=0,1$.
\end{enumerate}
%The positivity and unconfoundedness assumptions here condition on $\mathbf{X}_{hk},\mathbf{V}_h$ and $U_h$, but not $C_{hk}$. 
%This means \textit{there are cases where $\tau$ is identified but $\tau_h$ may be not}, for instance, when positivity with respect to $(\mathbf{X}_{hk},\mathbf{V}_h,U_h)$ is met in the population but is not met within clusters. Fortunately, our target estimand is $\tau$, not $\tau_h$.
%
Our challenge is that identification of $\tau$ requires $U_h$. Estimators that ignore this unobserved covariate are biased. \yl{Throughout this paper, we assume that we observe all of the individual-level covariates $\mathbf{X}_{hk}$ and some cluster-level covariates $\mathbf{V}_{h}$, but we do not observe $U_{h}$.}

\yl{For our proposed method to be effective in reducing bias from $U_{h}$, we need assumptions (a)-(c) as well as an assumption of an additive effect of confounding and modification due to any form of $U_{h}$ on $Y(0)$ and $Y(1)$. No distributional assumptions regarding the treatment assignment are required. More details are discussed in Section~\ref{sec:method}.}

%To simplify presentation, we also assume $Y_{hk}(z),Z_{hk}\independent C_{hk}\mid\mathbf{X}_{hk},\mathbf{V}_h,U_h$, that is, $\mathbf{V}_h$ and $U_h$ capture all the influence of the cluster's characteristics on both treatment assignment and potential outcomes. Not assuming this does not change the problem, but complicates notation and obscures arguments.

%% 
\section{Motivating application}
\label{sec:motivation}
As an illustrative example, consider an analysis of the ECLS-K (1998-1999) data~\citep{tourangeau2009early} mentioned in the Introduction. The data is publicly available at \url{https://nces.ed.gov/ecls/}. This is a longitudinal cohort with children nested in schools in the U.S. 
It has been widely used to study the effect of early education~\citep{hong2005effects, carlson2008physical, adelson2012examining} and behavioral interventions~\citep{xu2006does, gershoff2018strengthening} on child development. We consider the effect of center-based pre-school education on math proficiency as one illustrative example. Here center-based pre-school education means a child's exposure to any center-based childcare before kindergarten. There has been literature that reveals positive associations between exposure to center-based education and child development. Often these associational findings have an impact on government investments in pre-school programs~\citep{elango2015early}. Therefore, it is essential to fully understand their impacts on young children.
However, whether these associations imply causal effects is still intangible~\citep{loeb2007much, gottfried2015can}.

In this example, $Z_{hk}$ indicates whether child $k$ in school $h$ had primarily attended center-based pre-school ($Z_{hk} = 1$) or primarily received parental care ($Z_{hk} = 0$) before kindergarten, and $Y_{hk}$ is the child's observed math score in the fall of kindergarten \yl{(generally around age 5 in the United States)}. We restrict the sample (i) to the 778 schools with at least one treated child and one control child, and (ii) to the children with complete data on a set of observed covariates. These covariates (e.g., demographics, family characteristics, census region, etc.) are considered confounders that should be adjusted for, and \yl{we describe them in detail in the supplementary material}. This analysis takes the sample as the inference population; that is, our goal is to estimate the ATE for the children in the sample.

\begin{table}
\caption{\label{tab:region} \yl{Average (standard deviation) of} treatment and outcome given cluster-specific characteristic: census region.}
\centering
\resizebox{0.5\textwidth}{!}{\begin{tabular}{lrr}
  \hline
Census region & Treatment prevalence &  Average outcome \\ 
  \hline
   Midwest & 0.73 (0.45) &  28.88 (9.48) \\ 
 Northeast & 0.71 (0.45) &  27.97  (9.75)\\ 
  South & 0.69 (0.46) &  26.64 (9.20)\\ 
 West & 0.60 (0.49) &  25.98 (9.25)\\ 
   \hline
\end{tabular}}
\end{table}

\begin{figure}
\centering
\includegraphics[width=0.48\textwidth]{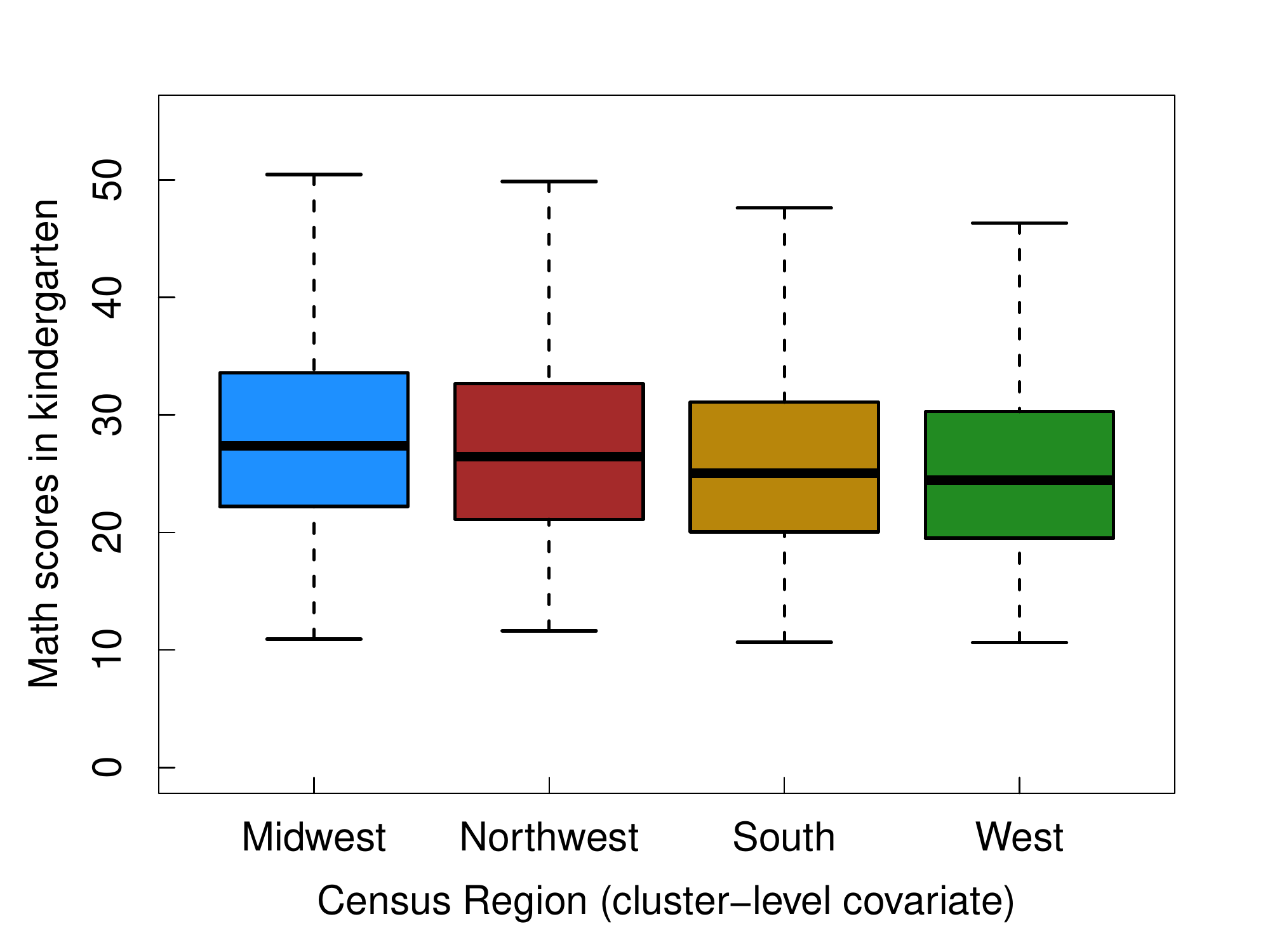}
\includegraphics[width=0.48\textwidth]{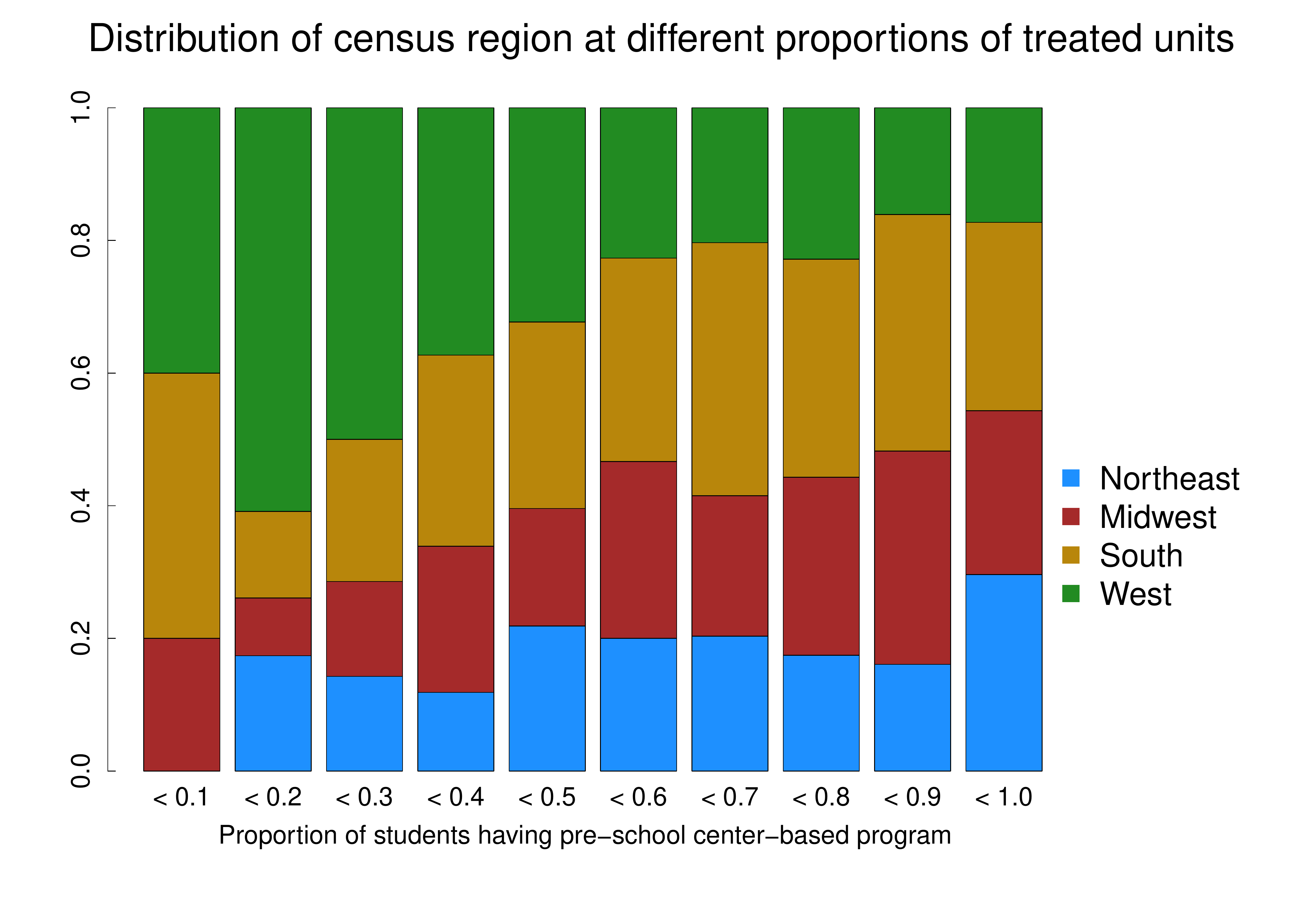}
\caption[]{\yl{The left panel shows the distribution of outcomes per each census region (cluster-level covariate) and the right panel shows the}
distribution of census region among schools in each decile of treatment prevalence.}
\label{fig:example}
\end{figure}

Within each cluster, individuals share the same value of each cluster-level variable. Hence a cluster-level confounder \yl{affects the treatment assignment probability (and the outcomes) for all individuals in the cluster with the same observed individual-level characteristics} in the same way. Consequently, cluster-level confounders are generally associated with the cluster's treatment prevalence and average outcome. 
In our example, we observe these two associations in the variable ``census region". Table \ref{tab:region} shows that census regions with higher treatment prevalence are also those with higher average outcome \yl{(see the left panel of Figure~\ref{fig:example})}. \yl{This suggests that there are contextual factors associated with census region that are confounders or effect modifiers}.
%; thus, census region might act as a proxy for these confounders.} 

%While it is insensible to think of census region as ``causing'' pre-school education or math proficiency, one could imagine various economic, cultural and other contextual factors (associated with census region) that may affect pre-school supply and demand and influence children's math proficiency. That is, census region is a proxy for contextual factors (mostly unmeasured) that may confound the causal effect of interest. 
%These factors may vary state-to-state, county-to-county, school-district-to-school-district, and school-to-school, so the four-category variable census region is a rather coarse proxy. While we can adjust for census region, the more important goal is to seek a tool that effectively deals with the unmeasured contextual factors, i.e., $U_h$. Such a tool would also be helpful where no proxy is available.

%Intuitively, there is some information on the unobserved $U_h$ in the observed cluster-specific treatment prevalence, $p_h:=\sum_{k=1}^{n_h}Z_{hk}/n_h$, which we can leverage. 
\yl{Even though the census regions are often known in survey data, let us \textit{hypothetically} presume that we do not observe this variable and treat is as $U_{h}$.}
The tendency of cluster-level confounders \yl{(e.g., census region)} to be associated with treatment prevalence provides a hint for a solution. 
\yl{The right panel of} Figure~\ref{fig:example} illustrates the relationship between census region and treatment prevalence among schools stratified by each school's treatment prevalence. For each stratum, the distribution of census region varies. Of the schools with treatment prevalence under 30\% (i.e., less than 30\% of the children had attended center-based pre-school), half were located in the West. Among schools with treatment prevalence over 80\%, however, less than one fifth were in the West, while most were in the South or the Midwest. 
This implies that schools that are similar in treatment prevalence are more likely to be located in the same census region\yl{, i.e., similar with respect to \textit{unobserved} cluster-level confounders}, than schools that are distant in this respect.

Building on this insight, our proposed method pools schools into several groups so that within each group the schools have similar observed treatment prevalence. With this grouping scheme, schools with similar $U_h$ values are more likely to be grouped together than schools with distant $U_h$ values. This is expected to reduce bias in the causal estimate due to the omitted $U_h$ through the propensity scores that at least partially take into account the unobserved $U_h$.

\section{Related Work}
\label{sec:review}

There are two stages of using propensity score methods to estimate causal effects in non-experimental settings:  estimation and use. The focus of the current paper is propensity score estimation. Of the different ways the estimated propensity scores can be used in estimating the ATE, we only examine inverse probability weighting (IPW), to focus attention on \yl{the} benefit of improved propensity score estimation as one example of the use of propensity scores in outcome analysis.
%. The estimators we investigate do not involve modeling the outcome; this is intended to shed light on the performance of the estimators as a result of improved propensity score estimation. 
Given this scope of the investigation, we review relevant strategies for propensity score estimation and briefly introduce IPW estimation in a multilevel setting, followed by a summary of existing methods designed specifically to handle unmeasured cluster-level confounding.

\subsection{Propensity score estimation with clustered data}
%As mentioned earlier, \textit{cluster-specific propensity score estimation} is not a realistic option for clusters with small sample sizes. Hence we do not pay much attention for this estimation strategy here, except to highlight a point that will become relevant later, that cluster-specific propensity score estimation requires positivity with respect to $\mathbf{X}_{hk}$ to hold within clusters: given any cluster $h$, for all $\mathbf{x}$ in the support of $\mathbf{X}_{hk}$ in the cluster, $0<p(Z_{hk}=1\mid\mathbf{X}_{hk}=\mathbf{x},C_{hk}=h)<1$. Note that this is a stronger assumption than the unconditional positivity in Section \ref{sec:settingetc}.

\yl{Propensity score estimation with fixed and/or random effects} has been \yl{often} used in practice in multilevel settings, as compared to \textit{cluster-specific propensity score estimation}; this is mostly due to small cluster size and stringent identification conditions within clusters, e.g., cluster-specific propensity score estimation requires $\mathbf{Z}_{h} = (Z_{h,1}, \ldots, Z_{h, n_{h}}) \neq \mathbf{0}, \mathbf{1}$ for all clusters.
Instead, \yl{multilevel} propensity score models incorporate the clustered structure with fixed or random cluster effects, thus allowing some degree of model heterogeneity across clusters. Many studies have shown that adding fixed or random effects in a propensity score model and/or outcome model improves performance~\citep{hong2006evaluating, thoemmes2011use, arpino2011specification, li2013propensity}. In our context with unobserved $U_h$, a propensity score model with fixed or random intercepts seems suitable, as the varying intercepts help absorb the effect of $U_h$ on treatment assignment. \yl{Therefore, if the cluster sizes are large enough, using a cluster fixed effects model might be the most reasonable approach.} 

However, these models still have practical challenges. 
A fixed effects model requires a large number of parameters ($H-1$ parameters for any association, e.g., intercept or slope coefficient, that is allowed to vary across the clusters). This makes the model unstable or not identified if clusters are very small, as there may not be enough information to estimate cluster-specific parameters \citep{li2013propensity}. This is the same problem with cluster-specific propensity score estimation, just to a lesser degree.
Random effects models use fewer parameters than fixed effects models, and cluster-specific effects for small clusters simply collapse to the population mean.
However, a random effects model requires a distributional assumption for each cluster-varying parameter, and this assumption may not be correct.
Perhaps more importantly, a random effects model assumes no correlation between the predictors in the model and the random effects, that is, no correlation between observed and unobserved covariates~\citep{skinner2011inverse, li2013propensity, he2018inverse, yang2018propensity}, an assumption that is unlikely to hold in practice.

\subsection{Outcome analysis for the ATE}

This discussion of the outcome analysis presumes that propensity scores have been estimated based on the whole observed data with random effects; \yl{we refer to this propensity score estimation as (full-RE)}, which is standard practice for most of propensity score-based estimators \yl{with a small cluster size}.  
We focus our outcome analysis on the IPW estimation, but this discussion of how much locally the treatment effects are measured first can be extended to other estimators, e.g., doubly robust estimators, that use the propensity scores.
Denote the treatment probability of the individual by $e_{hk}:=p(Z_{hk}=1\mid \mathbf{X}_{hk},\mathbf{V}_h,U_h)$. 
% By the assumptions in Section \ref{sec:settingetc}, $e_{hk}$ is also equal to $p(Z_{hk}=1\mid \mathbf{X}_{hk},\mathbf{V}_h,U_h)$ and $p(Z_{hk}=1\mid \mathbf{X}_{hk},C_{hk})$.
Denote the estimated propensity score by $\hat e_{hk(\text{\yl{full-RE}})}$. IPW estimation is based on the inverse probability of assigned treatment weight, $w_{hk}=Z_{hk}e_{hk}^{-1}+(1-Z_{hk})(1-e_{hk})^{-1}$, in this case estimated by $\hat w_{hk(\text{\yl{full-RE}})}=Z_{hk}{\hat e_{hk(\text{\yl{full-RE}})}}^{-1}+(1-Z_{hk})(1-\hat e_{hk(\text{\yl{full-RE}})})^{-1}$.

We discuss two such estimators. The first is the simple IPW estimator that is agnostic to whether the setting is multilevel. It involves weighting each individual by $\hat w_{hk(\text{\yl{full-RE}})}$, and taking the difference between the weighted mean outcome of treated individuals and the weighted mean outcome of untreated individuals to estimate $\tau$. Formally,
\begin{equation}
    \label{eq:tau-fullfull}
    \hat{\tau}_{\mbox{\scriptsize{(\yl{full-RE}, full)}}} :=  \frac{\sum\limits_{h=1}^{H} \sum\limits_{k=1}^{n_{h}} Z_{hk} \hat{w}_{hk(\text{full-RE})}  Y_{hk}}{ \sum\limits_{h=1}^{H} \sum\limits_{k=1}^{n_{h}} Z_{hk} \hat{w}_{hk(\text{full-RE})}} - \frac{\sum\limits_{h=1}^{H} \sum\limits_{k=1}^{n_{h}} (1-Z_{hk}) \hat{w}_{hk(\text{full-RE})} Y_{hk}}{\sum\limits_{h=1}^{H} \sum\limits_{k=1}^{n_{h}} (1-Z_{hk}) \hat{w}_{hk(\text{full-RE})}}.
\end{equation}
We index this estimator (and all other estimators of $\tau$) by double subscripts: the first component indicates the propensity score estimation strategy, and the second indicates the IPW strategy for ATE estimation given the estimated propensity score weights. $\hat{\tau}_{\mbox{\scriptsize{(full-RE, full)}}}$ involves fully pooled propensity score estimation with random effects (intercepts), followed by fully pooled (or marginal) IPW. The latter labeling is appropriate, as the weighted averaging of the outcomes pools all treated individuals and all untreated individuals over the full sample. 
A marginal IPW would result in consistent estimation of $\tau$ had the estimated weights been consistent for the correct weights $w_{hk}$ under the identification assumptions; but the problem here is that the estimated propensity scores and weights,  $\hat e_{hk\text{(full-RE)}}$ and $\hat w_{hk\text{(full-RE)}}$, do not fully take into account the unobserved $U_h$, leading to a biased estimate $\hat\tau_{(\text{full-RE, full})}$. 

Instead of marginal IPW, in the multilevel setting, \cite{li2013propensity} suggest a cluster-weighted IPW estimator that combines estimates of cluster ATEs ($\tau_h :=\mathbb{E}[\tau_{jk}\mid C_{jk}=h]$). Using the same propensity scores estimated via fully pooled models with random effects, $\tau_h$ is estimated for each cluster:
\begin{equation}
\label{eq:tauh-full}
    \hat{\tau}_{h\text{(full-RE)}} :=  \frac{ \sum\limits_{k=1}^{n_{h}} Z_{hk} \hat{w}_{hk\text{(full-RE)}} Y_{hk} }{ \sum\limits_{k=1}^{n_{h}} Z_{hk} \hat{w}_{hk\text{(full-RE)}} } -  \frac{\sum\limits_{k=1}^{n_{h}} (1-Z_{hk}) \hat{w}_{hk\text{(full-RE)}} Y_{hk}}{\sum\limits_{k=1}^{n_{h}} (1-Z_{hk}) \hat{w}_{hk\text{(full-RE)}}}.
\end{equation}
Then these cluster-specific effects are averaged over the clusters to estimate $\tau$, weighted by the sum of the propensity score weights for each cluster, $\hat w_{h\text{(full-RE)}}:=\sum_{k=1}^{n_h}\hat w_{hk\text{(full-RE)}}$,
\begin{equation}
    \label{eq:tau-fullcluster}
    \hat{\tau}_{\mbox{\scriptsize{(\yl{full-RE}, cluster)}}} := \frac{ \sum\limits_{h=1}^{H}  \hat{w}_{h\text{(full-RE)}} \hat{\tau}_{h{\mbox{\scriptsize{(full-RE)}}}} }{ \sum\limits_{h=1}^{H}\hat{w}_{h\text{(full-RE)}} }.
\end{equation}
The second subscript of $\hat\tau_\text{(\yl{full-RE}, cluster)}$ reflects the use of cluster-specific IPW.
This estimator requires that each cluster has at least one treated and one untreated individual, that is $\mathbf{Z}_{h} = (Z_{h,1}, \ldots, Z_{h, n_{h}}) \neq \mathbf{0}, \mathbf{1}$ for all clusters $h = 1,2,\ldots, H$. Note that this requirement is specific to this estimator, not a condition required for identification of $\tau$.

Despite additional conditions for estimation, there are several benefits of $\hat\tau_\text{(\yl{full-RE}, cluster)}$ for our unobserved $U_h$ setting. We found in our simulations that without effect modification, i.e., when treatment effects are homogeneous across the clusters, $\hat{\tau}_{\text{(\yl{full-RE}, cluster)}}$ is protective against confounding due to unmeasured cluster-level characteristics. This is because the estimation of cluster-specific ATE does not suffer from imbalance in $U_{h}$.
%cluster-specific ATE $\tau_{h}$'s are estimated first, which rules out the possibility of over- or under-representation of treatment or control group from particular clusters.
However, the relative weights among clusters, $\{\hat{w}_{h \text{(full-RE)}} : h=1,2,\ldots,H\}$, are still biased.
Under effect modification where $\tau_{h}$ varies across clusters, the errors in the estimated cluster weights $\hat{w}_{h, \text{(full-RE)}}$ due to missing $U_{h}$ in the propensity score model might over- or under-estimate the relative influence of $\tau_{h}$ in evaluating the overall ATE $\tau$, which leads to bias in ATE even if each of $\tau_{h}$'s were correctly estimated.

\subsection{Propensity score methods for unmeasured cluster-level confounding}
\label{ssec:others}

\cite{yang2018propensity} and \cite{he2018inverse} proposed propensity score methods that incorporate additional information related to $U_h$, thus improving upon the estimator in (\ref{eq:tau-fullfull}), \yl{under the same assumption we make that there are unobserved cluster-level confounders but no unobserved individual-level confounders}. 

\cite{yang2018propensity} proposes a calibration strategy that uses propensity score weights that satisfy the following conditions: (i) for each observed covariate, the weighted sum in each treatment arm equals the unweighted marginal sum, and (ii) in each cluster, the treated individuals' weights and the untreated individuals' weights both sum to $n_h$. Using these weights results in pseudo treated and untreated populations that replicate the means of observed covariates and of $U_h$ in the inference population. Consistent $\tau$ estimation using these weights is contingent on the assumption that both the true treatment assignment and outcome models are generalized linear models.

\cite{he2018inverse}, on the other hand, proposes conditional propensity score estimation based on sufficient statistics. Under certain conditions, \cite{he2018inverse} shows that conditioning on some function of the cluster's treatment assignment vector (in addition to the observed covariates) is sufficient to guarantee ignorability in the presence of unobserved $U_h$. Assuming a logit treatment assignment model, one such sufficient statistic turns out to be the number of treated individuals in the cluster, $s_{h} := \sum_{k=1}^{n_{h}} Z_{hk}$. \cite{he2018inverse} estimates propensity scores via maximum likelihood conditional on this statistic.
% Let $\hat{e}_{hk\text{(He)}}: = \hat p\left( Z_{hk} = 1 \mid \mathbf{X}_{hk}, \mathbf{V}_{h},C_{hk}, \mathcal{S}_{h} \right)$ denote propensity scores estimated conditional on this statistic. 
Marginal IPW based on the estimated conditional propensity scores is shown to reduce bias due to the unmeasured $U_h$.

\section{Methods}
\label{sec:method}

We now describe our proposed propensity score estimation method in some detail, and point out how it relates to existing methods. We then present two IPW estimators based on the estimated propensity scores, and discuss their theoretical properties.

\subsection{Selective pooling of cluster groups}
\label{ssec:grouping}
Our method relies on pooling information. This idea in fact has been used for similar purposes to ours. For example, \cite{stuart2008matching}, \cite{arpino2016propensity}, and  \cite{zubizarreta2017optimal} each provide strategies to match individuals across clusters to increase comparability in individual-level covariates. In our context, instead of pooling information from all the clusters indiscriminately, we selectively group similar clusters to guarantee large enough samples for propensity score estimation. This method therefore overcomes the small cluster sample size problem in the sense that it allows more variation in propensity score models than the alternative of a fully pooled model in this case. Group-stratified propensity score estimation requires positivity to hold within cluster groups, which is stricter than unconditional positivity, but less strict than the positivity within clusters required by cluster-specific propensity score estimation.

As mentioned in Section \ref{sec:motivation}, our selective pooling is based on the cluster's treatment prevalence, $p_h:=\sum_{k=1}^{n_h}Z_{hk}/n_h$, leveraging $U_h$'s association with treatment. Within the groups that are pooled, the variance of $U_h$ is likely smaller than its marginal variance, which means group-specific propensity score models are less misspecified than a fully pooled model. This method essentially taps into the same information about $U_h$ as \cite{yang2018propensity} and \cite{he2018inverse}: for each cluster, the number of treated individuals (used by \cite{he2018inverse}) carries the same information as the cluster's treatment prevalence; and \cite{yang2018propensity}'s tying the sums of the cluster's treated and control weights to the cluster sample size is another way of using that same information. Unlike this prior work, our proposed method is not based on assumptions regarding specific treatment assignment or outcome models.

To update notation, let $H_g$ denote the set of clusters that are grouped into group $g$, ($g=1,\dots,G$, $G\leq H$). Let $\hat e_{hk\text{(group)}}$ denote the propensity score estimated for individual $k$ in cluster $h$ based on the propensity score model \yl{using all of the observed individual- and cluster-level covariates (e.g., $\mathbf{X}_{hk}, \mathbf{V}_{h}$)} for the group that cluster $h$ is grouped in. Let $\hat w_{hk\text{(group)}}:=Z_{hk} \hat e_{hk\text{(group)}}^{-1}+(1-Z_{hk})(1-\hat e_{hk\text{(group)}})^{-1}$ denote the inverse probability weight of the individual based on $\hat e_{hk\text{(group)}}$.

%In implementation, the analyst needs to make two decisions: how to select the groups, and how to estimate propensity scores within groups. 
%There are many options for group selection, and we do not venture to make specific recommendations. How many groups to use, for example, could be a topic for investigation (similar to the question how many subclasses in a propensity score subclassification analysis); the answer may depend on the number of clusters and the distribution of cluster treatment prevalence. Here we simply choose $G=10$; in the data example this means roughly 78 clusters per group. 
There are many options for group selection that form multiple groups with similarity. 
In our case, to select $G$ groups out of $H$ clusters to minimize within-group distances in $p_h$, we use Partitioning Around Medoids (PAM) \citep{van2003new,park2009simple}. Other clustering methods, such as $k$-mean clustering \citep{hartigan1979algorithm} \yl{or the Bayesian nonparametric clustering~\citep{teh2005sharing}}, may also be applied. We will shortly discuss advantages of minimizing within-group distances in $p_{h}$'s.  

\subsection{Two IPW estimators based on the partially pooled propensity scores}

\begin{figure}
	\begin{centering}
	\resizebox{\textwidth}{!}{%
		\begin{tikzpicture}[>=stealth, node distance=1cm, every node/.style={rectangle,rounded corners, minimum width=0.5cm, minimum height=0.5cm, draw=black, fill=white!30, draw}]
		\tikzstyle{square} = [thin, rectangle, draw]
		\begin{scope}[xshift=-2.0cm]
		\path[->, thin]
	
		node[yshift=0cm,  line width=0.3mm, text width=16.5cm, align=center] (1) {(1) Compute the observed treatment prevalence for each cluster, $p_h$}
		node[below of=1,line width=0.3mm, text width=16.5cm, align=center] (2) {(2) Select cluster groups to minimize within-group distances in $p_h$}
		node[below of=2,  line width=0.3mm, text width=16.5cm, align=center] (3) {(3) Fit a propensity score model for each group \yl{with all observed covariates and random effects} and estimate propensity scores}

		%node[ below of=3,  line width=0.3mm, xshift = -2cm, text width=5cm, yshift = 0.5cm, draw = none] (3a) {Enough cluster size}
		
		%node[ below of=3,  line width=0.3mm, xshift = 6cm, text width=5cm, yshift = 0.5cm, draw = none] (3b) {Not enough cluster size}
		
		node[ below of=3,  line width=0.3mm, xshift = -4.5cm, text width=8cm, yshift = -0.5cm] (4a) {(4.a) Estimate cluster-specific ATEs ($\tau_{h}$)}
		
		node[ below of=3,  line width=0.3mm, xshift = 4.5cm, text width=8cm, yshift = -0.5cm] (4b) {(4.b) Estimate group-specific ATEs ($\tau_{g}$)}
		node[ below of = 4b,  line width=0.3mm, text width = 8cm, yshift = -0.5cm] (5b) {(5.b) Combine the groups' $\tau_g$ estimates to estimate the population ATE ($\tau$)}
		
		node[ below of = 4a,  line width=0.3mm, text width = 8cm, yshift = -0.5cm] (5a) {(5.a) Combine the clusters' $\tau_h$ estimates to estimate the population ATE ($\tau$)}
		
		node[below of = 5b,  line width=0.3mm, text width = 8cm, yshift = -0.5cm, text centered, draw = none] (6b) {(b) (group-RE, group) estimator}
		
		node[ below of = 5a,  line width=0.3mm, text width = 8cm, yshift = -0.5cm, text centered, draw = none] (6a) {(a) (group-RE, cluster) estimator}
	
		(1) edge[thick, black] (2)
		(2) edge[thick, black] (3)
		
		(3) edge[thick, black] (4a)
		(3) edge[thick, black] (4b)
		
		(4a) edge[thick, black] (5a)
		(4b) edge[thick, black] (5b)
		;
		\end{scope}				
		\end{tikzpicture}%
	}
		\par\end{centering}
	\caption{\label{fig:flowchart} Flowchart for IPW estimation of population ATE using partially pooled propensity scores.}
\end{figure}
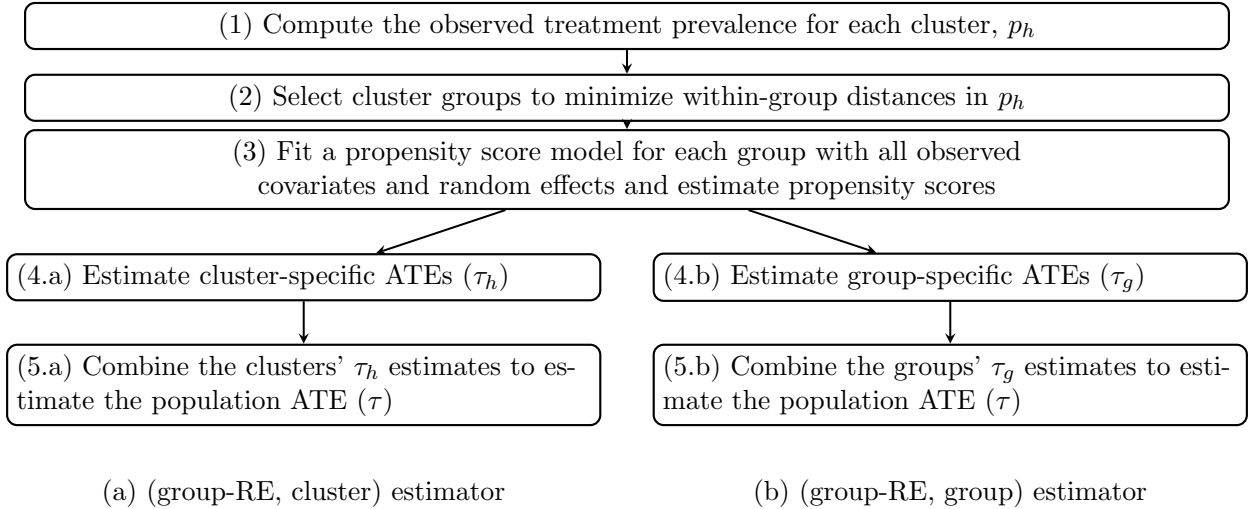

\yl{Given that we estimate the propensity scores after pooling clusters into several groups with similar treatment prevalence,}
we consider a cluster-weighted and a group-weighted estimator, denoted by $\hat\tau_\text{(group, cluster)}$ and $\hat\tau_\text{(group, group)}$, respectively. \yl{When we include cluster random effects in the propensity score models, we denote these cluster-weighted and group-weighted estimators by $\hat\tau_\text{(group-RE, cluster)}$ and $\hat\tau_\text{(group-RE, group)}$, respectively. By using a cluster-weighted (or group-weighted) IPW estimator, we can reduce the impact of unmeasured $U_{h}$ in causal effect estimation -- especially when $U_{h}$ modifies causal effects -- by \textit{locally} estimating the treatment effects from each cluster (or group of clusters with similar $p_{h}$) first and marginalizing them over clusters (groups) to estimate the ATE.}
Figure \ref{fig:flowchart} provides a skeleton summary of these two estimators.

The cluster-weighted estimator $\hat\tau_\text{(group-RE, cluster)}$ is a modification of $\hat\tau_\text{(full-RE, cluster)}$ replacing fully pooled weights $\hat{w}_{hk(\text{full-RE})}$ with group pooled weights $\hat{w}_{hk(\text{group-RE})}$, \yl{both with random effects}.
The group-weighted estimator $\hat{\tau}_\text{(group-RE, group)}$ using $\hat{w}_{hk(\text{group-RE})}$ is instead based on IPW at the group level.
\begin{equation}
    \label{eq:tau-groupgroup}
    \hat\tau_\text{(group-RE, group)}:=\frac{\sum\limits_{g=1}^G \hat w_{g\text{(group-RE)}} \hat\tau_{g\text{(group-RE)}}}{\sum\limits_{g=1}^G \hat w_{g\text{(group-RE)}}},
\end{equation}
where
\begin{equation}
    \label{eq:taug-group}
    \hat{\tau}_{g\text{(group-RE)}} :=  \frac{ \sum\limits_{h\in H_g} \sum\limits_{k=1}^{n_{h}} Z_{hk} \hat{w}_{hk\text{(group-RE)}} Y_{hk} }{\sum\limits_{h\in H_g}  \sum\limits_{k=1}^{n_{h}} Z_{hk} \hat{w}_{hk\text{(group-RE)}} } -  \frac{\sum\limits_{h\in H_g} \sum\limits_{k=1}^{n_{h}} (1-Z_{hk}) \hat{w}_{hk\text{(group-RE)}} Y_{hk}}{\sum\limits_{h\in H_g} \sum\limits_{k=1}^{n_{h}} (1-Z_{hk}) \hat{w}_{hk\text{(group-RE)}}}
\end{equation}
estimates the ATE for group $g$, $\tau_g:=\mathbb{E}[\tau_{hk}\mid C_{hk}\in H_g]$; and the group weight $\hat w_{g\text{(group)}}$ is the sum of the weights of all the individuals in the group, $\hat w_{g\text{(group-RE)}}:=\sum_{h\in H_g}\sum_{k=1}^{n_h}\hat w_{hk\text{(group-RE)}}=\sum_{h\in H_g}\hat w_{h\text{(group-RE)}}$. This estimator requires at least one treated and one untreated individual in each group, which is a milder condition required than for $\hat{\tau}_{\text{(group-RE, cluster)}}$ (one treated and one untreated each cluster). In practice, even though the $\hat{\tau}_{\text{(group-RE, cluster)}}$ estimator is less biased than the $\hat{\tau}_{\text{(group-RE, group)}}$ estimator, the $\hat{\tau}_{\text{(group-RE, cluster)}}$ may not be identified with small cluster size; we will discuss this issue further in Section~\ref{ssec:ipw_choice}.
We now turn to examine more closely the advantages of minimizing within-group dissimilarity in treatment prevalence.

\subsection{Decomposition of IPW estimator}
\label{ssec:theory}

We further elaborate the derivations for $\hat{\tau}_{\text{(group, group)}}$ under minimal distributional assumptions on outcomes to demonstrate why selectively pooling cluster groups with respect to similar treatment prevalence can reduce bias due to unmeasured context. 
We also make similar arguments about $\hat{\tau}_\text{(group, cluster)}$ in the supplementary material.

Assume a continuous outcome model where an arbitrary function of unmeasured cluster-level covariates has an additive effect on the outcome. Let us ignore observed covariates $\mathbf{X}_{hk}$ and $\mathbf{V}_{h}$ for simplicity. 
Consider the following data generating model for potential outcomes $\{ Y_{hk}(0), Y_{hk}(1)\}$ having $U_{h}$ as a confounder and an effect modifier \yl{via some arbitrary functions $g(U_{h})$ and $f(U_{h})$ that we do not put any restriction on other than they are bounded}. Note that we do not require any assumption on the distribution of treatment assignment here.
\begin{equation}
\label{eq:simple}
    Y_{hk}(z) =   \beta_{0} +  g(U_{h}) +  \kappa z +  f(U_{h}) z + \epsilon_{hk},~\epsilon_{hk}\overset{i.i.d.}{\sim} (\mu_{\epsilon} = 0, \sigma^{2}_{\epsilon} < \infty),
\end{equation}
%\comment{[For ease of understanding, you could point out in words that U is an effect modifier.]}
for $z \in \{0,1\}$.
Then we can represent $\tau$ using two potential outcomes for each individual:
%\comment{[This representation does not require that Y(0) and Y(1) are observed. $\tau$ is a function of Y(0) and Y(1), period :-), whether we observe these guys or not.]}
\begin{eqnarray*}
    \tau & = & \mathbb{E}\left[Y_{hk}(1) - Y_{hk}(0) \right] \\ 
    & = & \kappa + n^{-1} \sum\limits_{h=1}^{H} n_{h} f( U_{h}).
\end{eqnarray*}
%\comment{[the second line in the equation doesn't look right. The E[] in the first line (by definition of tau) mean the average of $Y_{hk}(1)-Y_{hk}(0)$ over $h,k$, right? Then why is there that E[] in the second line?]}

Now consider a grouping method that $H$ clusters were classified into $G~(\leq H)$ groups. 
Let $p_{g}$ denote the treatment prevalence in cluster group $g~(g=1,2,\ldots, G)$; and define $\delta_{h} :=  p_{h} -  p_{g}$ for $h \in H_{g}$ as the deviation of each cluster's treatment prevalence from the average treatment prevalence of the group that the cluster belongs to. 
Note that pooling the clusters with similar treatment prevalence is equivalent to selectively pooling clusters to minimize the $\delta_{h}$. 
%\comment{[What do you mean by the word ``remind'' here?]} 

We can demonstrate that a grouping method that minimizes $\delta_{h}$ reduces bias from $U_{h}$ in ATE estimation.
Under a simple scenario where no observed covariates are correlated with $U_{h}$,
consider the estimated weights $\hat{w}_{hk \text{(group)}}$ from the partially pooled propensity score model that does not include $U_{h}$ and ignores clusters.
Then we have the following decomposition of $\hat{\tau}_{\text{(group, group)}}$: 
\begin{equation}
\begin{split}
\label{eq:bias1}
    \hat{\tau}_{\mbox{\scriptsize{(group, group)}}} &= \left( \sum\limits_{g=1}^{G} \hat{w}_{g} \right)^{-1} \sum\limits_{g = 1}^{G} \hat{w}_{g} \hat{\tau}_{g}  \\
    & \approx  \tau + \Lambda + \Delta,
\end{split}
\end{equation}
where
\begin{equation}
\label{eq:lambda}
\Lambda = n^{-1} \sum\limits_{h=1}^{H} n_{h} \delta_{h}(p^{-1}_{g} + (1-p_{g})^{-1})  g\left( U_{h} \right) +  n_{h} \delta_{h} p^{-1}_{g} f\left(U_{h}\right)
\end{equation}
denotes the bias introduced by an unmeasured $U_{h}$, and $\Delta$ is a weighted sum of random errors $\epsilon_{hk}$'s.
\yl{In the presence of observed confounders $\mathbf{X}_{hk}$ and $\mathbf{V}_{h}$ that are included in the propensity score model, the expectation of the residual term $\Delta$ converges to zero; this is because the estimated propensity scores are essentially the conditional expectation of $Z_{hk}$ given $\mathbf{X}_{hk}$ and $\mathbf{V}_{h}$.}
In Equation~\eqref{eq:bias1} we have $\approx$ instead of $=$ because we approximate $\hat{w}_{hk \text{(group)}}$ only through the treatment prevalence in each group as a proxy for partially pooled propensity scores.
Details about the derivation procedures as well as the remainder $\Delta$ are provided in the supplementary material.

The above decomposition of $\hat{\tau}_{\text{(group, group)}}$ implies that the IPW estimator is a consistent estimator for $\tau$ when both $\Lambda$ and $\Delta$ converge to zero as the number of clusters increases. Convergence of $\Lambda$ is much more demanding as it denotes a systematic bias due to unmeasured context while convergence of error terms in $\Delta$ directly comes from the assumption of random errors $\epsilon_{hk} \overset{i.i.d.}{\sim} (0, \sigma^2_{\epsilon})$. 
More specifically, given that $f(U_{h})$ and $g(U_{h})$ are bounded, $\Lambda$
involves two terms, $n^{-1}\sum\limits_{h=1}^{H} n_{h}\delta_{h} / p_{g}$ and $n^{-1}\sum\limits_{h=1}^{H} n_{h}\delta_{h} / (1-p_{g})$. 
%\comment{[I don't understand this sentence, perhaps because of the unusual use of ``should''. Do you mean a combination of two points: (i) the approximation requires that these two quantities converge to zero and (ii) this convergence happens when $U_h$ is bounded? Also, is this convergence as $H\rightarrow\infty$?]} 
Under some regularity assumptions, to have $\Lambda \rightarrow 0$ as $H \rightarrow \infty$, $H^{-1}\sum\limits_{h=1}^{H} \delta_{h} / p_{g}$ and $H^{-1}\sum\limits_{h=1}^{H} \delta_{h} / (1-p_{g})$  need to go to zero.
Since $p_{g}$ and $(1-p_{g})$ are both positive, smaller $\delta_{h}$ leads to smaller bias $\Lambda$ at fixed values of $p_{g}$ and $(1-p_{g})$.
We now discuss the bias term $\Lambda$ in more detail.

\subsection{Discussion on bias}
\label{ssec:bias}

%\comment{[It seems unnecessarily complicated to start with the outcome model assumption (5.3) and then switch to (5.5). Is it simpler to just include g(U) and f(U) from the begining, i.e., in (5.3)? I think that would be the best.
%If I am wrong here and it would indeed be more complicated to do it that way, then we should keep this two steps organization, but at the begining of section 5.3, about when stating model (5.3), you could tell the reader that you are starting with a simple model (for simplicity of presentation) but will consider a more flexible model at a later step. That would be better.]}

%\comment{[It would be cleaner to remove $\beta_1$ and $\kappa_2$ from equation (5.5).]}

%\comment{[Do we need $\epsilon_{hk}$ to be normal? Same for equation (5.3).]} 

%Then bias $\Lambda$ that is dependent on $U_{h}$ under partially pooled propensity scores, i.e., $\hat{e}_{hk} = n_{g,1} /n_{g}$, in a group-weighted IPW is as follows:
%\begin{equation}
%\label{eq:Lambda1}
%    \Lambda = n^{-1} \sum\limits_{h=1}^{H} n_{h} \delta_{h}(p^{-1}_{g} + (1-p_{g})^{-1}) \beta_{1} g\left( U_{h} \right) +  n_{h} \delta_{h} p^{-1}_{g}  \kappa_{2} f\left(U_{h} \right).
%\end{equation}
Bias $\Lambda$ in Equation~\eqref{eq:lambda} shows that grouping clusters with similar $p_{h}$'s reduces bias from $f(U_{h})$ and $g(U_{h})$ by forcing $\delta_{h}$ to be small, relatively smaller than $p_{g}$ and $1-p_{g}$. %\comment{[The phrase ``preferably smaller than pg and 1-pg'' is imprecise. Please consider revising or removing it.]} 
We can infer from the first term in $\Lambda$ that we need a smaller $\delta_{h}$ when the treatment prevalence of the group is either small or large (i.e., small $p_{g}$ or small $1-p_{g}$), and/or when the amount of confounding is large, i.e., large $|g(U_{h})|$; moreover, from the second term in $\Lambda$, it is clear that we need a smaller $\delta_{h}$ when the treatment prevalence of the group is small, i.e., small $p_{g}$, and/or when the amount of effect modification is large, i.e., large $|f(U_{h})|$. %\comment{[There are several instances of ``or/and'' in this paragraph and perhaps other places as well. In English, it is standard to use ``and/or'', so I suggest we use that too. Also, I think the first instance, pg or/and 1-pg, that one should simply be ``and''.]}

One caveat in partial pooling cluster groups is that smaller $\delta_{h}$, i.e., smaller deviation of a cluster's treatment prevalence from the group's treatment prevalence, might not compensate for extreme values of $p^{-1}_{g}$ or/and $(1-p_{g})^{-1}$. In fact grouping by similar treatment prevalence often leads to an extremely large value of these two inverse prevalences by grouping treatment-dominated clusters (thereby small $(1-p_{g})^{-1}$) or control-dominated clusters together (thereby small $p_{g}^{-1}$).
%\comment{[I think you mean small pg and (1-pg), without the inverse.]} 
Therefore, we might need a relatively narrower window of treatment prevalence, e.g., allowing particularly small $\delta_{h}$, in one group than others when $p_{g}$ is almost near zero or one.

The other thing to note is that $\Lambda$ (Equation~\eqref{eq:lambda}) is the estimated bias when the propensity score model does not adjust for $U_{h}$ in any way. This will not be true if a random effects model is fit. If a random intercept or slope for each cluster is included in the model, \yl{e.g., $\hat{\tau}_{(\text{group-RE}, \cdot)}$, instead of $\hat{\tau}_{(\text{group}, \cdot)}$, is used}, $\Lambda$ would over-estimate the actual bias. We may instead consider $U_{h}$ as the remaining unmeasured cluster-level characteristic after adjusting for random intercepts or slopes.
Then the bias is smaller, but the cluster grouping strategy proposed here will still help reduce bias.

\section{Numerical Experiment}
\label{sec:simulation}

Through simulations, we aim to explore (i) whether selectively pooling clusters with respect to similar $p_{h}$ can restrain the influence of unmeasured cluster-level characteristics more effectively than partial pooling by random selection or based on similar observed characteristics, and (ii) whether a combination of selectively pooled propensity scores and different types of IPW estimators performs better than those using full pooling in the presence of an unmeasured cluster-level confounder and/or effect modifier.

Simulation settings are as follows. Suppose that we have $H=200$ clusters with cluster size of $n_{h} \sim  \lfloor \mathcal{U}(5, 25) \rfloor$. %\comment{[``for convenience'' sounds not very good, it sounds like we are too casual. I recommend you explain the rationale for excluding clusters without individuals in both arm. I also think this is something that needs to be addressed earlier, perhaps in the Methods section, because it essentially is about the inference population, i.e., we would like it to include the full sample, but due to blah blah blah it is more reasonable to exclude certain clusters. So this is a point in the Methods section, because it is about what the proposed methods do and what they do not do. (By the way, I don't remember precisely whether the detail about tau(full/group, cluster) requiring at least one treated and one control in each cluster is still retained in this version or if it has been removed, but I think it is both convenient and coherent to keep it in the lit review and methods sections.) Then in the numerical experiment section, I recommend not mentioning excluding clusters at the point of describing the data generating model, but mentioning excluding clusters at the step that follows, the implementation of the analysis using the methods. This is more concordant with reality when we get data that already has those clusters (the data have been generated) and what we do with the data includes excluding those clusters.]} 
We assume a relatively small cluster size considering our research question and given common data situations.
We generate continuous observed covariates $X_{hk} \overset{i.i.d.}{\sim} \mathcal{N}(0,1)$ measured at the individual level and $V_{h} \overset{i.i.d.}{\sim} \mathcal{U}(-1,1)$ at the cluster level, and an unobserved cluster-level covariate $U_{h} \overset{i.i.d.}{\sim} \mathcal{U}(-2,2)$.
%\comment{[Before talking about generation of treatment and outcome, I'd mention the simulation of X, V and U. That can be brief, but at least should tell the reader a few things like these are continuous variables, whether they are independent or correlated, whether/how the distribution of X differs across clusters. Most of the time readers do not go to the supplements, and they may not need full information but they should have enough information to be able to judge/speculate for themselves whether the results are likely to be general or may be specific to a specific scenario we happen to use.]} 
Then the data generating models for treatment assignment and potential outcomes are as below:
\begin{equation}
\begin{split}
\label{eq:model}
\mbox{logit}(e^{*}_{hk}) & =  \alpha_{h,0} + \alpha_{1}\bar{X}_{h}  + \alpha_{2} (X_{hk} - \bar{X}_{h}) + \alpha_{3} V_{h}  + \alpha_{4}  (U_{h} - \bar{U});  \\
Y_{hk}(z)& =  \beta_{h,0} + \beta_{1}\bar{X}_{h} +  \beta_{2}(X_{hk} - \bar{X}_{h})  + \beta_{3} V_{h} + \beta_{4} (U_{h} - \bar{U})  \\ & + z  \left\{ \kappa_{h,0} + \kappa_{1} \bar{X}_{h} + \kappa_{2} (X_{hk} - \bar{X}_{h})  +  \kappa_{3} V_{h} + \kappa_{4} (U_{h} - \bar{U})^2  \right\}\\& + \mathcal{N}(0, \sigma_{y}).
\end{split}
\end{equation}
We further adjust treatment assignment model by taking $e_{hk}  =  0.7e^{*}_{hk} + 0.15$ to assure an adequate number of treated and control individuals within each cluster. In the real data analysis, clusters with all units assigned to one treatment arm were dropped to allow for easier comparison of the IPW estimators. 
%\comment{[Since the treatment assignment model is slightly complicated (including both the first line in (6.1) and the first sentence here, it's cleaner to present the treatment assignment model first, finish that, and then present the potential outcome model. And you are simplifying $/0.7^{-1}$, right? :-)]}
A set of intercepts $(\alpha_{h,0}, \beta_{h,0}, \kappa_{h,0} )$ represent cluster-level random effects; $\bar{X}_{h}$ and $(X_{hk} - \bar{X}_{h})$ represent the associations of individual characteristics as a form of an aggregated characteristic within cluster ($\bar{X}_{h}$) and an individual's relative difference within cluster ($X_{hk} - \bar{X}_{h}$) respectively. We allow for the square of $(U_{h} - \bar{U})$ as an effect moderator to examine the performance of each estimator under non-linear treatment effects with respect to $U_{h}$. \yl{A parameter set $(\kappa_{h,0}, \kappa_{1}, \kappa_{2}, \kappa_{3}, \kappa_{4})$ controls the effect heterogeneity through cluster random effects and observed and unobserved covariates.}
Detailed settings about these parameters can be found in the supplementary material.

For numerical experiments, we vary the parameter values of $(\alpha_{4}, \beta_{4}, \kappa_{4})$, which represent the influence of an unmeasured covariate $U_{h}$ on the treatment assignments ($\alpha_{4}$), outcome distributions ($\beta_{4}$), and treatment effects ($\kappa_{4}$), respectively. 
Non-zero $\alpha_{4}$ and $\beta_{4}$ implies the presence of confounding due to $U_{h}$ other than effect modification and non-zero $\kappa_{4}$ implies the presence \yl{of} effect modification by $U_{h}$. 
For propensity score estimation, we use random effects models (e.g., full-RE or group-RE) throughout the numerical experiments and application because they are commonly used to adjust for unmeasured heterogeneity~\citep{arpino2011specification, thoemmes2011use}.

\begin{figure}[ht]
	\centering
	\includegraphics[width=0.9\textwidth]{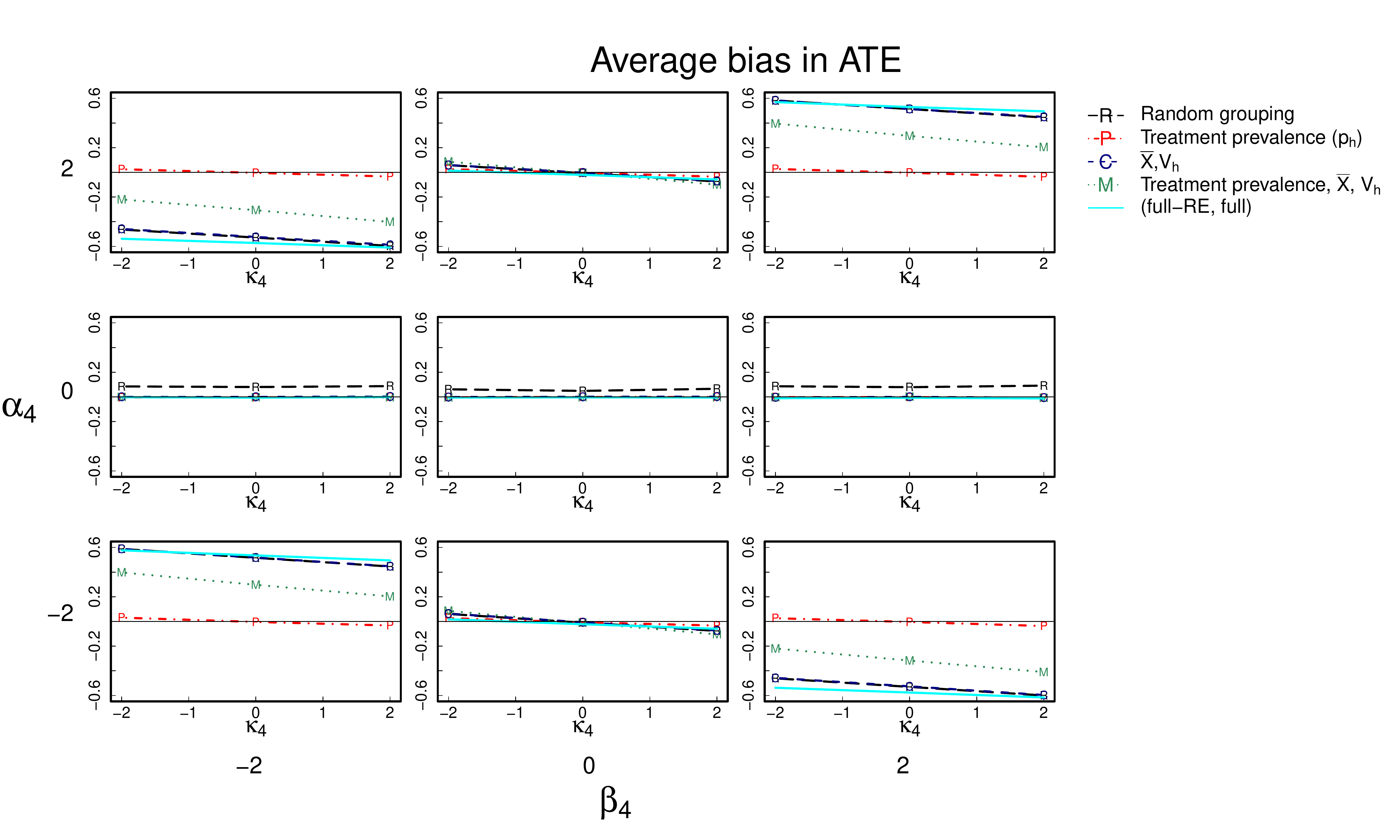}
\caption{\label{fig:sim112} Average bias in ATE over $r=1000$ replicates when different (group-RE, group) estimators as well as a (full-RE, full) estimator is used. For partially pooled propensity score models, $H=200$ clusters are selectively pooled into $G=10$ groups (R) Randomly, (P) by minimizing the within-group distance of the treatment prevalence within group, (C) by minimizing the within-group distance of the observed covariates, or (M) by minimizing the combination of the "merged" information including the cluster's treatment prevalence and the observed covariates.
Across over all scenarios, the (R), (C), and (full-RE, full) curves sit right on top of one another. \yl{After each partial pooling, propensity score models include all observed confounders ($X_{hk}$ and $V_{h}$) and random effects.}}
\end{figure}	

\subsection{Different choice of pooling methods}
Figure~\ref{fig:sim112} shows the average bias in the estimated ATE when different methods of cluster grouping are used given a fixed number of groups ($G=10$). After grouping we apply a $\hat{\tau}_{\text{(group-RE, group)}}$ estimator based on the partially pooled propensity scores estimated with observed covariates and random intercepts. Also, we illustrate the performance of a (full-RE, full) estimator (cyan lines), without any partial pooling. 
We observe 
that partially pooling cluster groups with similar treatment prevalence (red lines) results in the smallest average bias -- even smaller than that using additional information of observed covariates $\bar{X}_{h}$ and $V_{h}$ (green lines); whereas random grouping and grouping with observed covariates do not add any benefits compared to fully pooled propensity scores.

\begin{figure}[ht]
	\centering
	\includegraphics[width=0.8\textwidth]{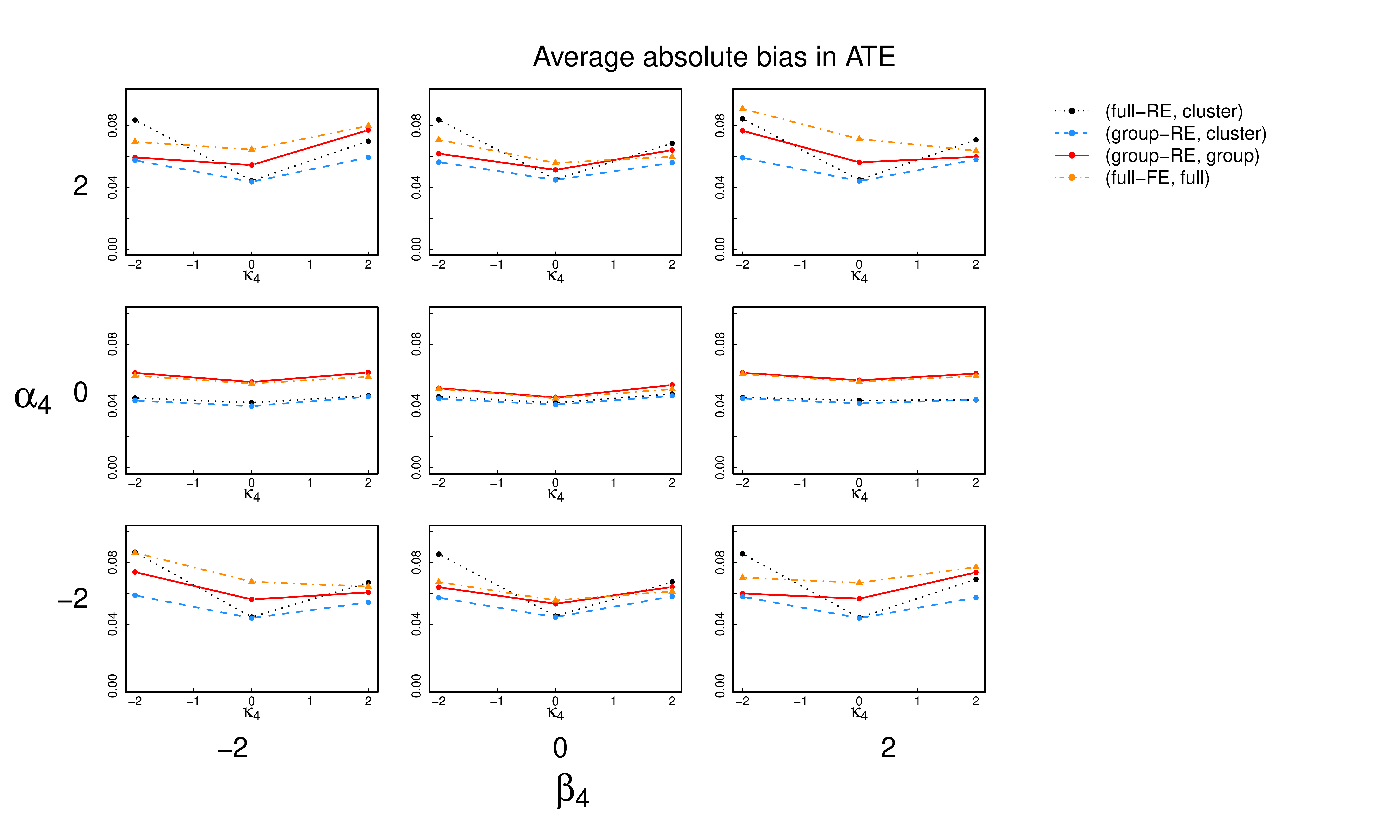}
\caption{\label{fig:absbias} Average absolute bias in ATE over $r=1000$ replicates when pooled propensity score model with random effects (\yl{full-RE}, $\cdot$) or partially pooled propensity score model of cluster groups with random effects (group-RE, $\cdot$) is used; as a causal estimator, a group-weighted ($\cdot$, group), and cluster-weighted ($\cdot$, cluster) IPW are applied following propensity score estimation. \yl{For comparison purposes, we also include (full-FE, full) which presents the results when propensity score models with fixed effects were used.}}
\end{figure}

\begin{table}
\caption{\label{tab:simtable}
\yl{Bias, absolute bias, standard error, and coverage rate of 95\% confidence intervals when $(\alpha_{4}, \beta_{4}, \kappa_{4}) = (-2, -2, -2)$.
 Standard deviation of the potential outcome under control is one.
(*) About 10\% of fixed effects models fail to converge; so we needed to drop the fixed effects terms. (**) On average, for about 10 out of $H=200$ clusters, $\tau_{h}$'s were not identified; in this case, the corresponding clusters were ignored in the ATE estimation.}
}
\centering
\resizebox{0.6\textwidth}{!}{\begin{tabular}{rrrrr}
  \hline
 & Bias & $|$Bias$|$ & SE & Coverage \\ 
  \hline
*(full-RE, full) & -0.039 & 0.086 & 0.509 & 1.000 \\ 
  (full-RE, full) & 0.577 & 0.577 & 0.134 & 0.001 \\ 
  (full-RE, group) & 0.119 & 0.127 & 0.142 & 0.971 \\ 
  (full-RE, cluster) & 0.079 & 0.087 & 0.094 & 0.938 \\ 
  (group-RE, full) & 0.014 & 0.079 & 0.426 & 0.999 \\ 
  (group-RE, group) & 0.000 & 0.074 & 0.146 & 1.000 \\ 
  **(group-RE, cluster) & 0.025 & 0.059 & 0.097 & 0.992 \\ 
   \hline
\end{tabular}}
\end{table}

\subsection{Different choice of IPW estimators and standard error estimation}
\label{ssec:ipw_choice}

Now only using the partial pooling with similar similar treatment prevalence, we compare our proposed IPW estimators -- (group-RE, cluster) and (group-RE, group) -- to a (full-RE, cluster) estimator. In Figure~\ref{fig:absbias}, we present the average of \textit{absolute} bias in ATE estimates under model~\ref{eq:model}.
When fully pooled propensity scores are used, a (full-RE, cluster) estimator reduces the influence from $U_{h}$ better compared to (full-RE, full); but it is still sensitive to effect modification due to $U_{h}$, i.e., under non-zero $\kappa_{4}$. 
Overall, a cluster-weighted IPW analytically and theoretically has smaller bias than group-weighted IPW \yl{as it estimates $\tau_{h}$ while being free from unmeasured confounding. However, the cluster-weighted IPW becomes invalid to use when the cluster size is so small that some clusters have all individuals in the same treatment condition}.

Standard errors assuming fixed propensity scores and coverage rates based on those are provided in the supplementary material \yl{Table S1}. 
\yl{Table~\ref{tab:simtable} above presents the simulation results under one of the scenarios when $(\alpha_{4}, \beta_{4}, \kappa_{4}) = (-2, -2, -2)$. Note that we could not always identify the ATE using the fixed effects or the cluster-weighted IPW estimator. Even though the fixed effects model substantially reduces bias compared to the random effects model, it produces large variability mainly due to the small cluster size.}
\yl{Moreover, in Table~\ref{tab:simtable}, it appears that a (group-RE, cluster) looks better than a (group-RE, group), but for about 10\% out of $r=1000$ replications, the (group-RE, cluster) estimator distorts the population as it discards the clusters of which $\tau_{h}$'s were not identified.}

\yl{In the supplementary material, we provide additional simulations to examine whether those combinations perform better than the existing methods proposed for unmeasured context problems~\citep{he2018inverse, yang2018propensity}. 
The results demonstrate that the partial pooling method is most robust against interactions between the observed and unobserved covariates.
We also compared the proposed methods to latent class approach~\citep{kim2015multilevel} which was not developed for the current purpose of handling unobserved cluster-level confounding, but also results in a grouping of clusters. With a continuous $U_{h}$, partial pooling by treatment prevalence $p_{h}$ performs better than partial pooling by the estimated latent class.}

\section{Application to ECLS-K data}
\label{sec:application}
%\core{data descriptions:}
We apply our partially pooled propensity score method to estimate 
%\comment{[identify or estimate?]} 
the causal effect of pre-school programs on children's math achievement in kindergarten. \yl{In this setting, one child's pre-school participation is not likely to affect the math achievement of other children in the same school; thus, we assume that SUTVA holds within and across clusters.}
We consider potential confounding due to child-level characteristics (sex, race, age, family type and motor skills), and school-level characteristics (census region (Northeast/Midwest/South/West) and location (Central City/large town/rural)) for the propensity score models.
%\comment{[in kindergarten, no ``the'', like in primary school, in high school]}. 
%This is the same question discussed in \textbf{[Dong et al.]} and we use the same set of individual-level and cluster-level covariates with those present at \textbf{[Dong et al]} except for socioeconomic status (SES), family income, and continuous parent education.
%To illustrate the role of clusters we deliberately do not include strong individual-level covariates such as socioeconomic status (SES), family income, and parent education.
Detailed information is in \cite{dong2020using} and the supplementary material.

%\comment{[This sounds a bit confusing. Do you mean to say that you are going to present results from (1) analyses that include these two cluster-level covariates as well as (2) analyses that pretend we don't have these variables (to amplify ``unobserved'' cluster-level confounding)? Related to this, I think it could also be said that for both cases -- observing or not observing these covariates -- there may be unobserved cluster-level confounding, which our two estimators address.]} 
We focus on $H=778$ schools with at least one student under treatment (center-based program) and under control (parental care) in the school to enable us to apply a cluster-weighted IPW.
These schools were then categorized into $G=10$ groups based on the treatment prevalence within school using the PAM method.

We have two observed cluster-level covariates -- census region and location. Accordingly, we consider two scenarios to explore the role of partially pooled propensity scores:
\begin{itemize}
    \item (Model 1) Propensity score models include these two cluster-level covariates as well as the observed individual-level covariates.
    \item (Model 2) Propensity score models do not include any cluster-level covariates but only include the observed individual-level covariates. 
\end{itemize}
The motivation behind considering Model 2 is to compare different estimators with potential unobserved cluster-level covariates that we indeed observed. Of course in both Models 1 and 2 there might also be other unobserved cluster-level characteristics, we expect Model 2 would suffer more from unobserved confounding  under the absence of two cluster-level covariates.
We also added a random effect to each propensity score model considering its wide use in multilevel data settings. %\comment{[Again, this justification is repetitive (see my comments in the simulations section). I recommend moving it to the Methods section as one of the options that can be used, with a comment on its advantage (additional protection again variation in U within clusters), which is better than just saying we use it because it is commonly used. I think I wrote this in the Methods section, which I hope you consider using.]}
In \yl{Table~\ref{tab:cov_balance}}, we provide the covariates' balance table of the two cluster-level covariates before weighting and after weighting, and weighting by fully and \yl{partially} pooled propensity scores under Model 2. The results show that even in the absence of important cluster-level covariates in the propensity score model, using partially pooled propensity scores helps to reduce bias from those unmeasured covariates\yl{, resulting in reduced standard mean differences for census region and location that had not been included in the propensity score model. Details of the balance on all of the observed covariates are presented in the supplementary material.}

\begin{table}
\caption{\label{tab:cov_balance} Covariate balance table for the two cluster-level covariates when observations are not weighted by propensity scores (unweighted PS), weighted by partially pooled PS, and weighted by fully pooled PS.  Both partially pooled propensity score model and fully pooled propensity score model omitted cluster-level variable of census region and location (Model 2) and included random effects.
}
\centering
\resizebox{\textwidth}{!}{\begin{tabular}{r||lll|lll|lll}
  \hline
  & \multicolumn{3}{c|}{Unweighted PS} 
  & \multicolumn{3}{c|}{Partially pooled PS} & 
  \multicolumn{3}{c}{Fully pooled PS} \\
Variable & Parental care & Center-based & SMD & Parental care & Center-based & SMD & Parental care & Center-based & SMD \\ 
  \hline
  \textbf{Census region}  &    &   &  0.211 &   &   &  0.026 &   &   &  0.080 \\ 
Northeast &    473 (17.5)  &  1165 (19.9)  &  &  1696.1 (19.9)  &  1620.2 (18.9)  &  &  1503.5 (19.1)  &  1613.3 (19.4)  &  \\ 
 Midwest &    549 (20.3)  &  1465 (25.0)  &  &  2025.6 (23.7)  &  2024.0 (23.6)  &  &  1761.0 (22.4)  &  2019.2 (24.3)  &  \\
  South &    894 (33.1)  &  2034 (34.7)  &  &  2852.1 (33.4)  &  2894.1 (33.8)  &  &  2632.4 (33.4)  &  2855.5 (34.4)  &  \\ 
   West &   784 (29.0)  &  1193 (20.4)  &  &  1963.0 (23.0)  &  2023.9 (23.6)  &  &  1978.6 (25.1)  &  1819.8 (21.9)  &  \\ 
 \textbf{Location}   &   &   &  0.190 &   &   &  0.041 &   &   &  0.124 \\ 
  Central city &   1110 (41.1)  &  2363 (40.3)  &  &  3401.0 (39.8)  &  3509.6 (41.0)  &  &  3118.6 (39.6)  &  3402.4 (41.0)  &  \\ 
Large town &    1000 (37.0)  &  2594 (44.3)  &  &  3552.7 (41.6)  &  3594.1 (42.0)  &  &  3137.3 (39.8)  &  3586.8 (43.2)  &  \\ 
  Small town &       590 (21.9)  &   900 (15.4)  &  &  1583.1 (18.5)  &  1458.5 (17.0)  &  &  1619.6 (20.6)  &  1318.5 (15.9)  &  \\ 
  Sum of weights & 2700 &  5857 &  & 8536.83 & 8562.19 &  & 7875.49 & 8307.70 &  \\ 
   \hline
\end{tabular}}
\end{table}

\begin{figure}
\centering
\includegraphics[width=0.9\textwidth]{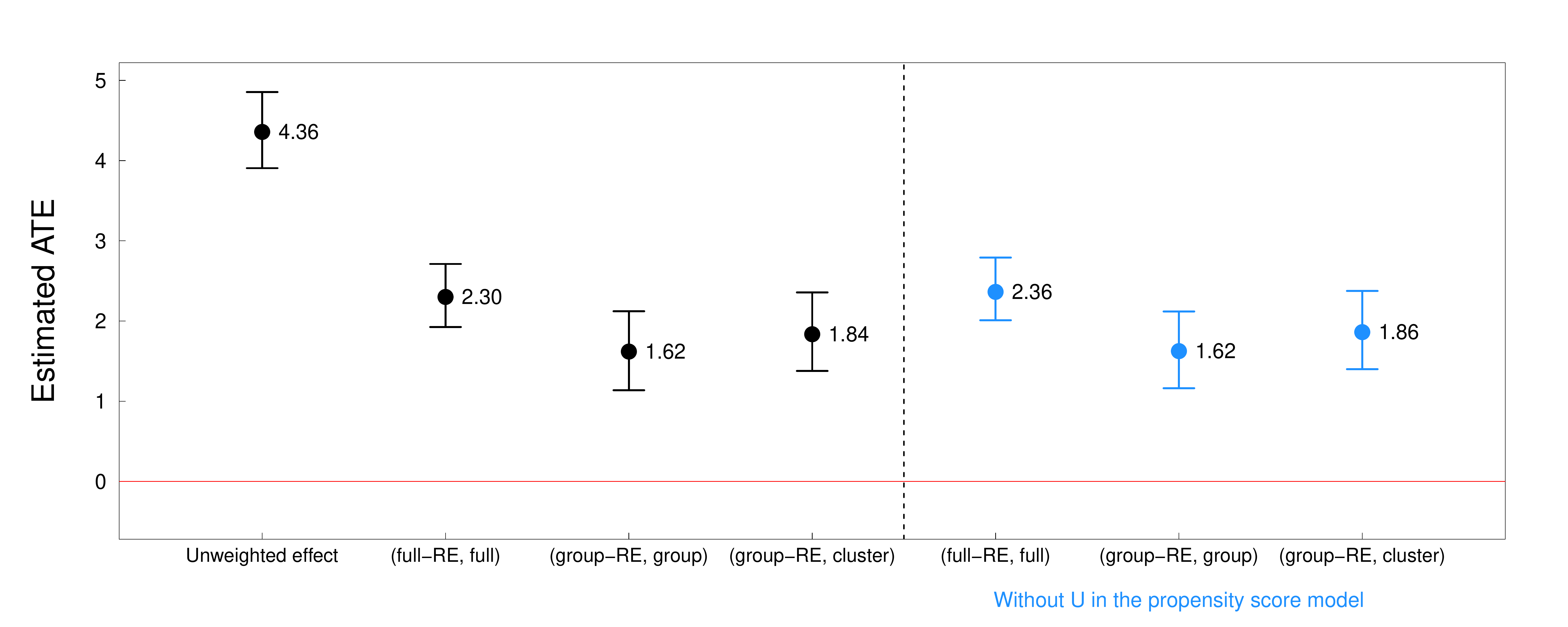}
\caption{\label{fig:real_result} Point estimates (dots) and empirical 95\% confidence intervals (vertical lines) for the estimated causal effect using (i) no propensity score weighting, (ii) a marginal IPW with pooled propensity scores, i.e., (full-RE, full), (iii) a group-specific IPW with partially pooled propensity scores, i.e., (group-RE, group), and (iv) a cluster-specific IPW with partially pooled propensity scores, i.e., (group-RE, cluster). 
The last three blue lines show the results of Model 2 under the (full-RE, full), (group-RE, group), and (group-RE, cluster) estimators when the propensity score models intentionally omitted two cluster-level characteristics: census region and location.}
\end{figure}

We summarize the results under different schemes in Figure~\ref{fig:real_result}. First of all, it is evident that the causal estimate (4.36) without weighting (Unweighted effect)
%\comment{[It is unclear here which estimator you are referring to. Is this a propensity score estimator? But not the (full, full)? If not a PS estimator, should we even consider it here?]} 
might overestimate the causal effect of center-based programs over parental care given lack of adjsutment for confounding, compared to any of the propensity score-weighted estimates. 
The following three black lines in Figure~\ref{fig:real_result} show the results under Model 1 using different methods.
We observed that when individuals were weighted by partially pooled propensity scores and causal effects were estimated through a group-weighted IPW, i.e., using the (group-RE, group) estimator (Equation~\eqref{eq:tau-groupgroup}), the size of effect of center-based program (1.62) is substantially lower than that under the combination of pooled propensity scores and a marginal IPW (2.30). 
%\comment{[Note that the word ``so-called'' has a bit of a negative connotation: it is called that, but it is not quite that. Example: ``so-called friends'' are those you call friends but may not be true friends.]}  
The result using a (group-RE, cluster) estimator (1.84) shows a slightly higher point estimate than that (1.62) using a (group-RE, group) estimation approach. \yl{The average and the standard deviation of observed math scores are 27.27 and 9.45, respectively. Therefore, the estimated effect size of $\hat{\tau}=1.84$ represents an increase of about $0.2$ standard deviations.}

%\core{noticeable observations:}
The results under Model 2 are presented in the last three blue lines. The results show that with partially pooled propensity score models, the estimates stay nearly constant even after missing census region and location, exhibiting slightly more robustness against missing two cluster-level covariates than (full-RE, full) estimator. In contrast, the fully pooled propensity scores still result in the estimates (2.36) closer to the unweighted effects than those under Model 1 (2.30).

%\comment{[I have a different thought about this that is related to my earlier comment about $p_h$ being associated with a function of $\{\bar X, V, U\}$. It may be easier to talk about this than to write something lengthy.]}

\section{Conclusion}
\label{sec:conc}

In this work, we discuss the use of a partially pooled propensity score estimation method to reduce bias in the causal estimate when unmeasured cluster-level characteristics influence treatment assignment and potential outcomes. We emphasize its usefulness when cluster sizes are small and the number of baseline characteristics is relatively large. We use simulation studies to examine the method's performance and apply the partially pooled propensity score approach to estimate the effect of pre-school education programs on children's performance.    %\comment{[I don't understand what you mean by ``provide a survey data of''.]}
%We restrict our inference population on the sample,
%but considering generalizing causal effect into super-population is often of main interest and of importance for policy perspectives in survey data~\citep{dugoff2014generalizing, austin2018propensity}, the immediate next step is to extend or adapt this method to evaluating population average treatment effect under complex survey data.  

Throughout simulations and applications, we only considered clusters having at least one treated and at least \yl{one} control individual in the sample. However, this is not a necessary condition to identify ATE ($\tau$) and the proposed method can allow clusters with treatment prevalence $p_{h} = 0$. Restricting the treatment prevalence is for our convenience to compare different types of causal estimators such as cluster-specific IPW (which requires $p_{h} > 0$) and group-specific IPW (which requires $p_{g} > 0$).  Still, it is important to note that restricting to clusters with $p_{h} > 0$ might distort the target population. \yl{For this reason, we can also benefit from the partial grouping of clusters by using the same group of clusters in the IPW estimator when a cluster-level IPW is not identified.}
%Therefore, it future work should mimic cluster-specific (or group-specific) outcome analysis (e.g. IPW estimators) while minimizing cluster-specific (group-specific) positivity assumptions.  
%\comment{[Reading this, I think we should be clear much earlier in the paper (in the Methods section where the two estimators are introduced) that the (group, cluster) estimator requires each cluster to have at least one treated and one control, but the (group, group) estimator does not. I think I made that clear in the version I edited, but this is a later version so I am not sure if it's still there. Please consider this.]}

We also found room for improvement in the method used to group clusters as we briefly discussed in Section~\ref{ssec:bias}. First of all, we observed that bias in an IPW estimator depends on the group's treatment prevalence ($p_{g}$) as well as on the difference between the cluster's treatment prevalence and the group's treatment prevalence ($\delta_{h}$); therefore, it would be a great help to develop data-adaptive grouping methods that could possibly vary the maximum of $\delta_{h}$ or vary the number of groups ($G$) depending on $p_{g}$. 
%\comment{[This point is great!!!]} 
\yl{Second, we will thoroughly investigate the changes in bias and variability as the number of groups changes, instead of fixing the number of groups as done here, e.g., $G=10$}. Moreover, as we have seen in the simulation studies a noticeable improvement in bias by using a cluster-weighted or group-weighted IPW, instead of using a marginal IPW, different grouping strategies in the \textit{use} of propensity scores as well as in the estimation can be further explored.
%\comment{[It is not very clear what this sentence is intended to say. Is this the point that in this paper we focused on grouping clusters to improve propensity score estimation, and different grouping strategies should be further explored for the next step of analysis, including estimation based on modeling the outcome?]}
%Lastly, we have assumed an untestable assumption of correctly specified individual-level characteristics. Sensitivity analysis of our method under its violation is important for practical purposes, which might vary depending on the correlation between unmeasured, individual-level and cluster-level characteristics. 

Overall, this partial pooling method provides straightforward and effective tools to reduce bias in causal effect estimation due to unmeasured contextual factors, and has many potential avenues for further development and application.

\section*{Acknowledgements}

This research was supported by the Institute of Education Sciences, U.S. Department of Education, through Grant R305D150001 awarded to Johns Hopkins University (PI's:  Stuart and Dong) and by Grant R01MH115487 from the National Institute of Mental Health (PI: Stuart). 
The opinions expressed are those of the authors and do not represent views of the Institute or the U.S. Department of Education. The authors are thankful to the members of the University of Pennsylvania Causal Inference Reading Group for helpful comments.

\bibliographystyle{Chicago}
\bibliography{reference}

\clearpage
\setcounter{equation}{0}
	\setcounter{figure}{0}
	\setcounter{table}{0}
	\setcounter{page}{1}
	\setcounter{section}{0}
	\renewcommand{\theequation}{S\arabic{equation}}
	\renewcommand{\thefigure}{S\arabic{figure}}
	\renewcommand{\thesection}{S\arabic{section}}
	\renewcommand{\thetable}{S\arabic{table}}
\renewcommand{\thefigure}{S\arabic{figure}}
\begin{center}
{\LARGE \textbf{Supplementary Material}}
\end{center}

\section{Asymptotic properties of ATE estimator}
\label{sec:asymptotic}
We consider the following data generating model for potential outcomes:
\begin{eqnarray*}
    Y_{hk}(z) & = & \beta_{0} + g(U_{h}) + \kappa z  +  z  f(U_{h})  + \epsilon_{hk},~\epsilon_{hk} \overset{i.i.d.}{\sim} (0, \sigma^{2}_{\epsilon}).
\end{eqnarray*}
Define $n_{h,z} = \sum\limits_{k=1}^{n_{h}} I\left( Z_{hk} = z \right)$ and~$n_{g,z} = \sum\limits_{h \in H_{g}} n_{h,z}$ for $z=0,1$, so that $p_{h} = n_{h,1}/n_{h}$.
Similarly, $p_{g} = n_{g,1}/n_{g}$ denotes a treatment prevalence of cluster group $g~(g = 1,2, \ldots, G)$. Then we have $\delta_{h} = n_{h,1} / n_{h} - n_{g,1} / n_{g} =  p_{h} -  p_{g}$ for $h \in H_{g}$.
With those notations, a (group, group) estimator, \yl{without random effects in the propensity score model}, can be represented as follows:

%\comment{[This is the supplement, which does not suffer from length limit, so could you put in below all the steps of the derivation? For example, it is not obvious how to connect from line 1 to line 2. I guess there are several steps that takes us from $\hat w_g$ to the various $n$s somehow, and also from $\hat\tau_g$ in eq. (5.2) to parameters of the $Y(z)$ model. I think from line 2 to line 3 also, there must be several steps in between.] }

\begin{eqnarray}
    \hat{\tau}_{\mbox{\scriptsize{(group, group)}}} &=& \left( \sum\limits_{g=1}^{G} \hat{w}_{g} \right)^{-1} \sum\limits_{g = 1}^{G} \hat{w}_{g} \hat{\tau}_{g} \nonumber \\
    & = & \left(\sum\limits_{g=1}^{n} 2n_{g}\right)^{-1} \sum\limits_{g=1}^{G} 2n_{g} \left( \frac{ \sum\limits_{h \in H_{g}}\sum\limits_{k=1}^{n_{h}} Z_{hk} n_{g,1} Y_{hk} / n_{g}}{ \sum\limits_{h \in H_{g}} \sum\limits_{k=1}^{n_{h}} Z_{hk} n_{g,1} / n_{g} } - \frac{ \sum\limits_{h \in H_{g}}\sum\limits_{k=1}^{n_{h}} (1- Z_{hk}) n_{g,0} Y_{hk} / n_{g} }{ \sum\limits_{h \in H_{g}} \sum\limits_{k=1}^{n_{h}} Z_{hk} n_{g,0} / n_{g}} \right) \nonumber \\
    &= & \kappa +  n^{-1} \sum\limits_{g=1}^{G} \sum\limits_{h \in H_{g} } \left\{ \frac{n_{g} n_{h,1}}{n_{g,1}}   f(U_{h}) + \left( \frac{n_{g} n_{h,1}}{n_{g,1}} - \frac{n_{g} n_{h,0}}{n_{g,0}} \right)  g(U_{h})  \right\}  + \Delta \nonumber \\ 
    & =&  \kappa + n^{-1}  \sum\limits_{h=1}^{H} n_{h}  f(U_{h})  \\ & \quad &  + n^{-1} \sum\limits_{h=1}^{H} n_{h}\left(\delta_{h} / p_{g}\right)  f(U_{h}) + n_{h} \delta_{h}(p^{-1}_{g} + (1-p_{g})^{-1})   g(U_{h})  + \Delta \nonumber \\ 
    & \approx & \kappa + n^{-1} \sum\limits_{h=1}^{H} n_{h}    f(U_{h})  + \Lambda \quad (\mbox{when } \Delta \rightarrow 0) \nonumber \\ 
    & \approx & \tau \quad (\mbox{when } \Lambda \approx 0), \nonumber
\end{eqnarray}
where
\begin{eqnarray}
\label{eq:delta}
    \Delta & = & n^{-1} \sum\limits_{g=1}^{G} n_{g} \left[\sum\limits_{h \in H_{g}} \left\{ \sum\limits_{k}^{Z_{hk} = 1} \frac{\epsilon_{hk}}{n_{g,1}} -  \sum\limits_{k}^{Z_{hk} = 0} \frac{\epsilon_{hk}}{n_{g,0}} \right\}  \right].
\end{eqnarray}
By the assumption of $\mathbb{E}(\epsilon_{h,k}) = 0$, $\Delta$ goes to $0$ as the number of clusters increases to infinity.

Here we used one trick: when the propensity scores are estimated for each cluster,  $\hat{e}_{hk}$ is approximately $n_{h,1} / n_{h}$; when the propensity scores are estimated \textit{within} the partially pooled cluster groups, they can be approximated by $\hat{e}_{hk, \text{(group)}} = n_{g,1} / n_{g} ;~h \in H_{g}, k = 1,2, \ldots, n_{h}$ without random intercepts. There might be some perturbations due to other covariates involved, e.g., $X_{hk}$ and $V_{h}$, but the expectation of propensity scores evaluated within group would be that group's treatment prevalence. These estimated propensity scores are analogous to those under propensity score model with fixed effects (e.g., when one dummy variable is spent per one cluster)~\citep{li2013propensity}.
Consequently, the second equation in \yl{Equation (8)} in the main manuscript comes from a simple derivation of $\hat{w}_{g \text{(group)}} = \sum\limits_{h,k,h \in H_{g}} Z_{hk} \hat{e}^{-1}_{hk \text{(group)}} + (1-Z_{hk}) (1-\hat{e}_{hk \text{(group)}})^{-1} = 2 n_{g}$, so $\sum\limits_{g=1}^{G} \hat{w}_{g} = 2 n$.
%\comment{[Oh I see part of the explanation here, but I think some steps are still implicit. Can you make all steps explicit?]}
We do not put any distribution assumptions for the treatment assignment $Z_{hk}$. If there is no unmeasured $U_{h}$ and all the other covariates are measured in the propensity score models, $\mathbb{E}[\delta_{h}] = 0$, i.e., $\Lambda$ would be close to zero.

\yl{In the presence of observed confounders $\mathbf{X}_{hk}$ and $\mathbf{V}_{h}$, consider $\epsilon_{hk}$ in Equation~\eqref{eq:delta} as $g(\mathbf{X}_{hk}, \mathbf{V}_{h}) + \epsilon^{\prime}_{hk}$ with random errors $\epsilon^{\prime}_{hk} \sim (0, \sigma^{2}_{\epsilon})$. Note that $\hat{e}_{hk} \approx n_{g,1}/n_{g}$.
Then the following equations show that the residual term $\Delta$ converges to zero as the number of clusters increases.
\begin{eqnarray*}
 \Delta & = & n^{-1} \sum\limits_{g=1}^{G} n_{g} \left[\sum\limits_{h \in H_{g}} \left\{ \sum\limits_{k}^{Z_{hk} = 1} \frac{\epsilon_{hk}}{n_{g,1}} -  \sum\limits_{k}^{Z_{hk} = 0} \frac{\epsilon_{hk}}{n_{g,0}} \right\}  \right] \\ 
 &\approx&  n^{-1} \sum\limits_{g=1}^{G}  \left[\sum\limits_{h \in H_{g}} \left\{ \sum\limits_{k}^{Z_{hk} = 1} \frac{\epsilon_{hk}}{\hat{e}_{hk}} -  \sum\limits_{k}^{Z_{hk} = 0} \frac{\epsilon_{hk}}{1-\hat{e}_{hk}} \right\}  \right] \\ 
&=& n^{-1} \sum\limits_{g=1}^{G}  \left[\sum\limits_{h \in H_{g}} \left\{ \sum\limits_{k}^{Z_{hk} = 1} \frac{g(\mathbf{X}_{hk}, \mathbf{V}_{h}) + \epsilon^{\prime}_{hk}}{E(Z_{hk} | \mathbf{X}_{hk}, \mathbf{V}_{h})} -  \sum\limits_{k}^{Z_{hk} = 0} \frac{g(\mathbf{X}_{hk}, \mathbf{V}_{h}) + \epsilon^{\prime}_{hk}}{E(1-Z_{hk} | \mathbf{X}_{hk}, \mathbf{V}_{h})} \right\}  \right]. 
\end{eqnarray*}
Therefore, the following equations hold.
\begin{eqnarray*}
 E(\Delta) 
&=& n^{-1} \sum\limits_{h=1}^{H} \sum\limits_{k=1}^{n_{h}}E  \left[   \frac{Z_{hk} \{g(\mathbf{X}_{hk}, \mathbf{V}_{h}) + \epsilon^{\prime}_{hk}\} }{E(Z_{hk} | \mathbf{X}_{hk}, \mathbf{V}_{h})} -   \frac{(1-Z_{hk})\{ g(\mathbf{X}_{hk}, \mathbf{V}_{h}) + \epsilon^{\prime}_{hk}\}}{E(1-Z_{hk} | \mathbf{X}_{hk}, \mathbf{V}_{h})}   \right] \\ 
&=& n^{-1} \sum\limits_{h=1}^{H} \sum\limits_{k=1}^{n_{h}}E  \left[  E \left\{  \frac{Z_{hk} \{g(\mathbf{X}_{hk}, \mathbf{V}_{h}) + \epsilon^{\prime}_{hk}\} }{E(Z_{hk} | \mathbf{X}_{hk}, \mathbf{V}_{h})} -   \frac{(1-Z_{hk})\{ g(\mathbf{X}_{hk}, \mathbf{V}_{h}) + \epsilon^{\prime}_{hk}\}}{E(1-Z_{hk} | \mathbf{X}_{hk},\mathbf{V}_{h})} \bigg| \mathbf{X}_{hk}, \mathbf{V}_{h} \right\}   \right]  \\ 
&=&  n^{-1} \sum\limits_{h=1}^{H} \sum\limits_{k=1}^{n_{h}}E  \left[   \frac{g(\mathbf{X}_{hk}, \mathbf{V}_{h})E ( Z_{hk} | \mathbf{X}_{hk}, \mathbf{V}_{h} ) + \epsilon^{\prime}_{hk} }{E(Z_{hk} | \mathbf{X}_{hk}, \mathbf{V}_{h})} -   \frac{g(\mathbf{X}_{hk}, \mathbf{V}_{h}) E(1-Z_{hk}| \mathbf{X}_{hk}, \mathbf{V}_{h})+ \epsilon^{\prime}_{hk}\}}{E(1-Z_{hk} | \mathbf{X}_{hk}, \mathbf{V}_{h})} \right]  \\ 
&=& n^{-1} \sum\limits_{h=1}^{H} \sum\limits_{k=1}^{n_{h}}E  \left[   \frac{g(\mathbf{X}_{hk}, \mathbf{V}_{h})E ( Z_{hk} | \mathbf{X}_{hk}, \mathbf{V}_{h} ) + \epsilon^{\prime}_{hk} }{E(Z_{hk} | \mathbf{X}_{hk}, \mathbf{V}_{h})} -   \frac{g(\mathbf{X}_{hk}, \mathbf{V}_{h}) E(1-Z_{hk}| \mathbf{X}_{hk}, \mathbf{V}_{h})+ \epsilon^{\prime}_{hk}\}}{E(1-Z_{hk} | \mathbf{X}_{hk}, \mathbf{V}_{h})} \right]  \\ 
&=& n^{-1} \sum\limits_{h=1}^{H} \sum\limits_{k=1}^{n_{h}}E  \left[   \frac{ \epsilon^{\prime}_{hk} }{E(Z_{hk} | \mathbf{X}_{hk}, \mathbf{V}_{h})} -   \frac{\epsilon^{\prime}_{hk}}{E(1-Z_{hk} | \mathbf{X}_{hk}, \mathbf{V}_{h})} \right]\\
&=& 0 .
\end{eqnarray*}
}

On the other hand, a (group, cluster) estimator can also be decomposed into the true effect $(\tau)$, bias due to $U_{h}$ ($\tilde{\Lambda}$), and a random error ($\tilde{\Delta}$) when $\hat{e}_{hk, \text{(group)}} = n_{g,1}/n_{g}$. 
\begin{eqnarray*}
\label{eq:bias2}
    \hat{\tau}_{\mbox{\scriptsize{(group, cluster)}}} & = & \left( \sum\limits_{h=1}^{H} \hat{w}_{h} \right)^{-1} \sum\limits_{h = 1}^{H} \hat{w}_{h} \hat{\tau}_{h}  \\
    & = & \kappa +  {(2n)}^{-1} \sum\limits_{h=1}^{H}   \left( \frac{n_{g} n_{h,1}}{n_{g,1}} + \frac{n_{g} n_{h,0}}{n_{g,0}} \right)  f(U_{h})  + \tilde{\Delta} \\
    & = & \kappa + n^{-1}  \sum\limits_{h=1}^{H} n_{h}  f(U_{h})  \\ & \quad &  + n^{-1} \sum\limits_{h=1}^{H}   n_{h} \delta_{h}(p^{-1}_{g} - (1-p_{g})^{-1}) f(U_{h})/2  + \tilde{\Delta} \\ 
    & \approx & \kappa + n^{-1} \sum\limits_{h=1}^{H} n_{h}   f(U_{h})  + \tilde{\Lambda} \quad (\mbox{when } \tilde{\Delta} \rightarrow 0)  \\ 
    & \approx & \tau \quad (\mbox{when } \tilde{\Lambda} \approx 0),
\end{eqnarray*}
Here biases due to $U_{h}$ and a random error are given by:
\begin{eqnarray*}
\tilde{\Lambda} & =&  n^{-1} \sum\limits_{h=1}^{H}   n_{h} \delta_{h}(p^{-1}_{g} - (1-p_{g})^{-1})   f(U_{h})/2, \\
\tilde{\Delta} &= &  n^{-1} \sum\limits_{h=1}^{H} \left( \frac{n_{h,1} n_{g}}{2 n_{g,1}} + \frac{n_{h,0} n_{g}}{2 n_{g,0}} \right) \left\{ \sum\limits_{k=1}^{Z_{hk} = 1} \epsilon_{hk}  / n_{h,1} - \sum\limits_{k=1}^{Z_{hk} = 0} \epsilon_{hk} / n_{h,0} \right\}.
\end{eqnarray*}

Note that under a (group, cluster) estimator, bias due to $g(U_{h})$ disappears and bias due to non-zero $f(U_{h})$ remains only. This supports our claim that a cluster-specific IPW is protective against confounding but still sensitive to effect modifications due to unmeasured confounders. 

%\comment{[This last point is an important point. Can you put it in the main text? I propose this for the main text: after discussing the (group, group) estimator, include a brief section on the (group, cluster) estimator to say that the bias only includes the component due to the effect modification.] }

\section{Simulation setting and additional results}
\label{sec:simresult}

Throughout the numerical experiments in this paper, we set $\alpha_{h,0}, \beta_{h,0}, \kappa_{h,0} \overset{i.i.d.}{\sim} \mathcal{N}(0,1)$ for $h=1,2,\ldots, H$; $(\alpha_{1}, \alpha_{2}, \alpha_{3}) = (-1, -1, -0.5)$; $(\beta_{1}, \beta_{2}, \beta_{3}) = (1, -1, 0.5)$; $(\kappa_{1}, \kappa_{2}, \kappa_{3}) = (-0.5, 1, -1)$; $X_{hk} \overset{i.i.d.}{\sim} \mathcal{N}(0,1)$; $V_{h} \overset{i.i.d.}{\sim} \mathcal{U}(-1,1); U_{h} \overset{i.i.d.}{\sim} \mathcal{U}(-2,2); \epsilon_{hk} \overset{i.i.d.}{\sim} \mathcal{N}(0, 1)$. When generating a set of covariates, we did not create any correlations among the variables. In practice, the covariates are often correlated with each other. Then $U_{h}$ indicates unmeasured cluster-level factors remaining after controlling $\bm{X}_{hk}$ and $\bm{V}_{h}$, such as the residuals in unmeasured cluster-level confounders after conditioning on all the observed covariates.

\begin{equation}
\begin{split}
\label{eq:general_outcome_other}
\mbox{logit}(e^{*}_{hk}) & =   \alpha_{h,0} + \alpha_{1} \bar{X}_{h}  + \alpha_{2}  (X_{hk} - \bar{X}_{h}) +   \alpha_{3} V_{h} +  \alpha_{4} (U_{h}-\bar{U})\\
e_{hk} & =   e^{*}_{hk}/0.7^{-1} + 0.15 \\
Y_{hk} & =   \beta_{h,0} +  \beta_{1} \bar{X}_{h}  +  \beta_{2} (X_{hk} - \bar{X}_{h}) +  \beta_{3} V_{h}  + \beta_{4}  (U_{h} - \bar{U})^{\beta^{\prime}}  + \\ & Z_{hk} \left\{ \kappa_{h,0} + \kappa_{1} \bar{X}_{h}  + \kappa_{2} (X_{hk} - \bar{X}_{h})  + \kappa_{3} V_{h}  +  (U_{h} - \bar{U})^{\kappa^{\prime}}  \kappa_{4} \right\}  + \epsilon_{hk}
\end{split}
\end{equation}
\yl{Figure~5} in the main text presents bias in $\hat{\tau}_{\text{(group-RE, group)}}$ (a modification of $\hat{\tau}_{\text{(group, group)}}$ with random intercepts in the propensity score model) when $( \beta^{\prime}, \kappa^{\prime}) = (1,2)$ with random intercepts in the propensity score models. Now assume that all of the information from $U_{h}$ has not been absorbed by random intercepts, i.e., use $\hat{\tau}_{\text{(group, group)}}$. Then bias due to $U_{h}$ is captured by following $\Lambda$.
\begin{equation}
\label{eq:approx_lambda}
    \Lambda \approx n^{-1} \sum\limits_{h=1}^{H} n_{h} \delta_{h}(p^{-1}_{g} + (1-p_{g})^{-1}) \beta_{4} (U_{h} - \bar{U})^{\beta^{\prime}} + n_{h} \delta_{h} p^{-1}_{g} \kappa_{4} (U_{h} - \bar{U})^{\kappa^{\prime}}  
\end{equation}
Different from \yl{Equation (9)} in the main text, here in Equation~\eqref{eq:approx_lambda}, we put $\approx$ instead of $=$ since now we have additional individual- and cluster-level covariates, $\bar{X}_{h}, (X_{hk}  - \bar{X}_{h})$, and $V_{h}$. Roughly speaking, here $\delta_{h}$ indicates the difference between the treatment prevalence of group and cluster that is not explained by the propensity score model. As additional covariates $(\bar{X}_{h}, (X_{hk}  - \bar{X}_{h}), V_{h})$ are all included in the propensity score model, we can represent the systematic bias only with respect to $U_{h}$. 
 
\begin{figure}[H]
	\centering
	\includegraphics[width=0.8\textwidth]{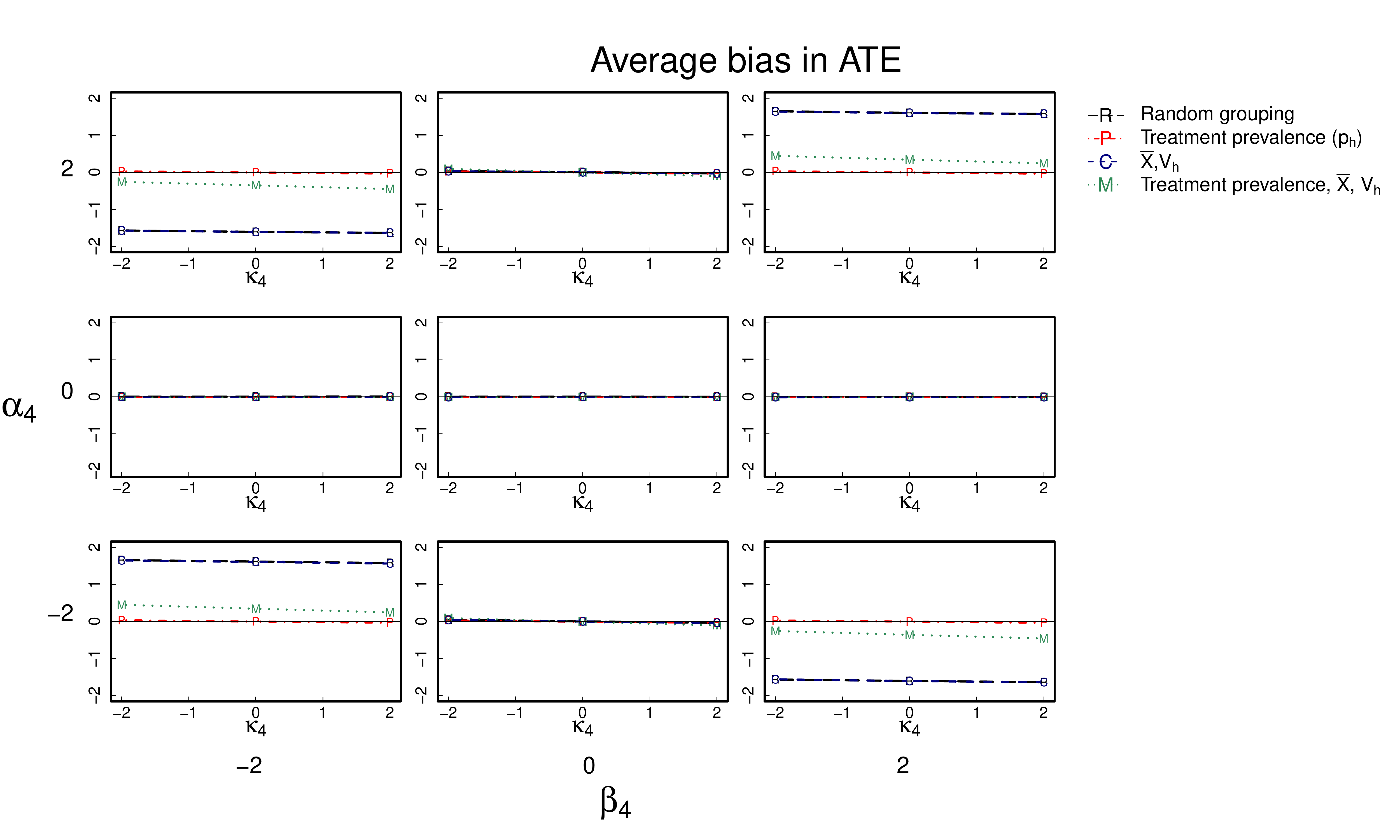} \\
	\includegraphics[width=0.8\textwidth]{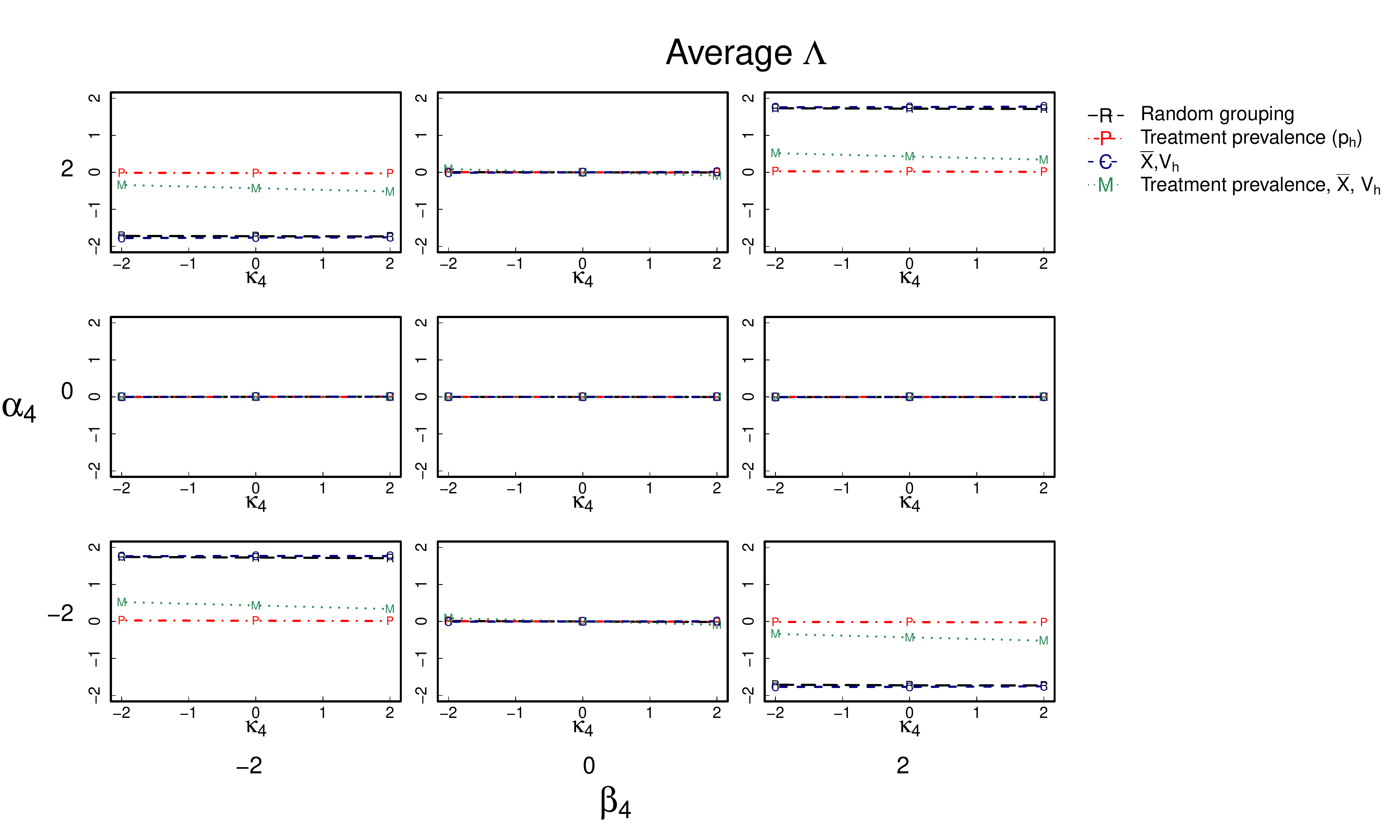} \\
\caption{\label{fig:sim112_norandom} Average bias in ATE (upper panel) and average estimated bias $\Lambda$ (lower panel) over $r=1000$ replicates when $(\beta^{\prime}, \kappa^{\prime}) = (1,2)$ \yl{when the propensity score models do not include random intercepts, i.e., $\hat{\tau}_{\text{(group, group)}}$ was used.}}	
\end{figure}	
 
\begin{figure}[H]
	\centering
	\includegraphics[width=0.8\textwidth]{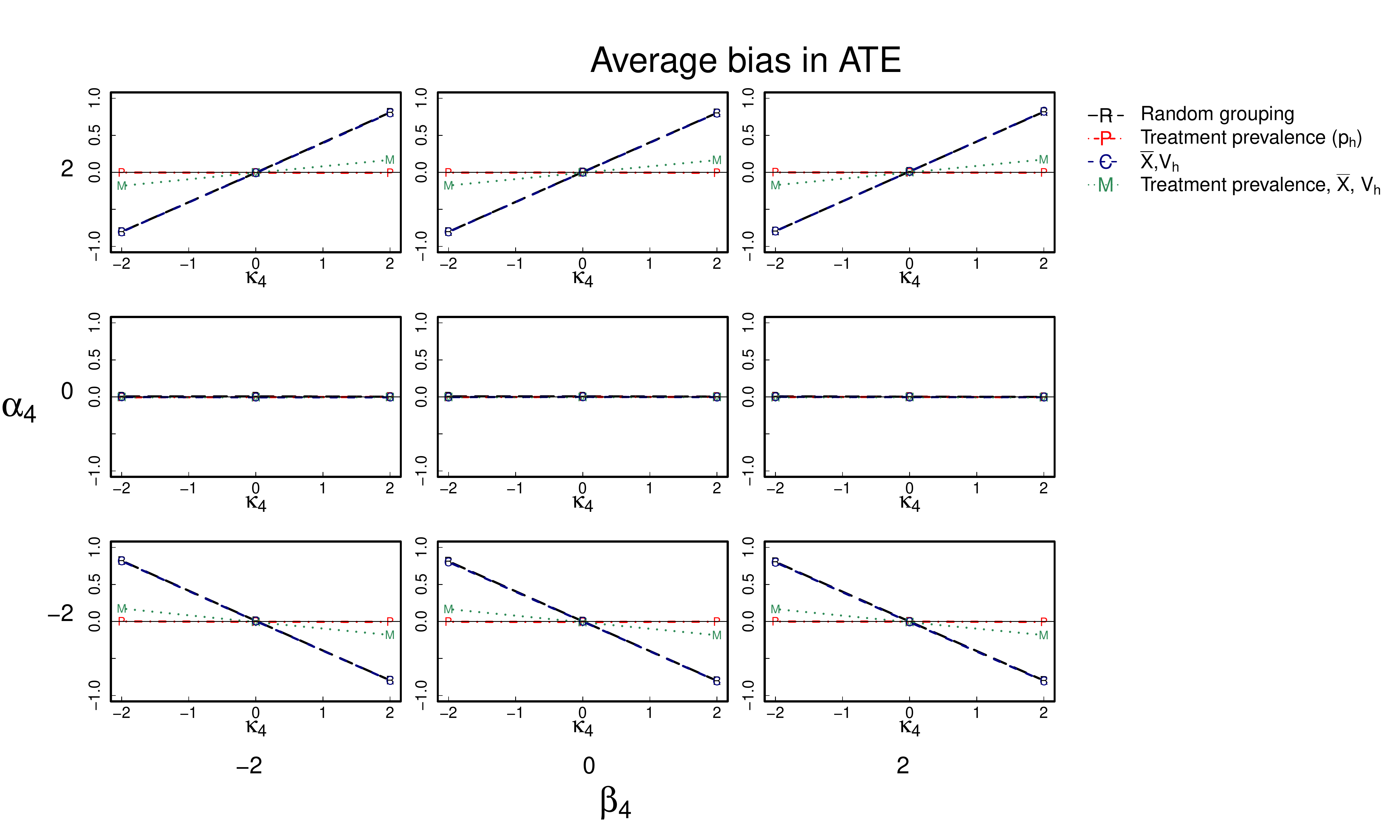} \\
	\includegraphics[width=0.8\textwidth]{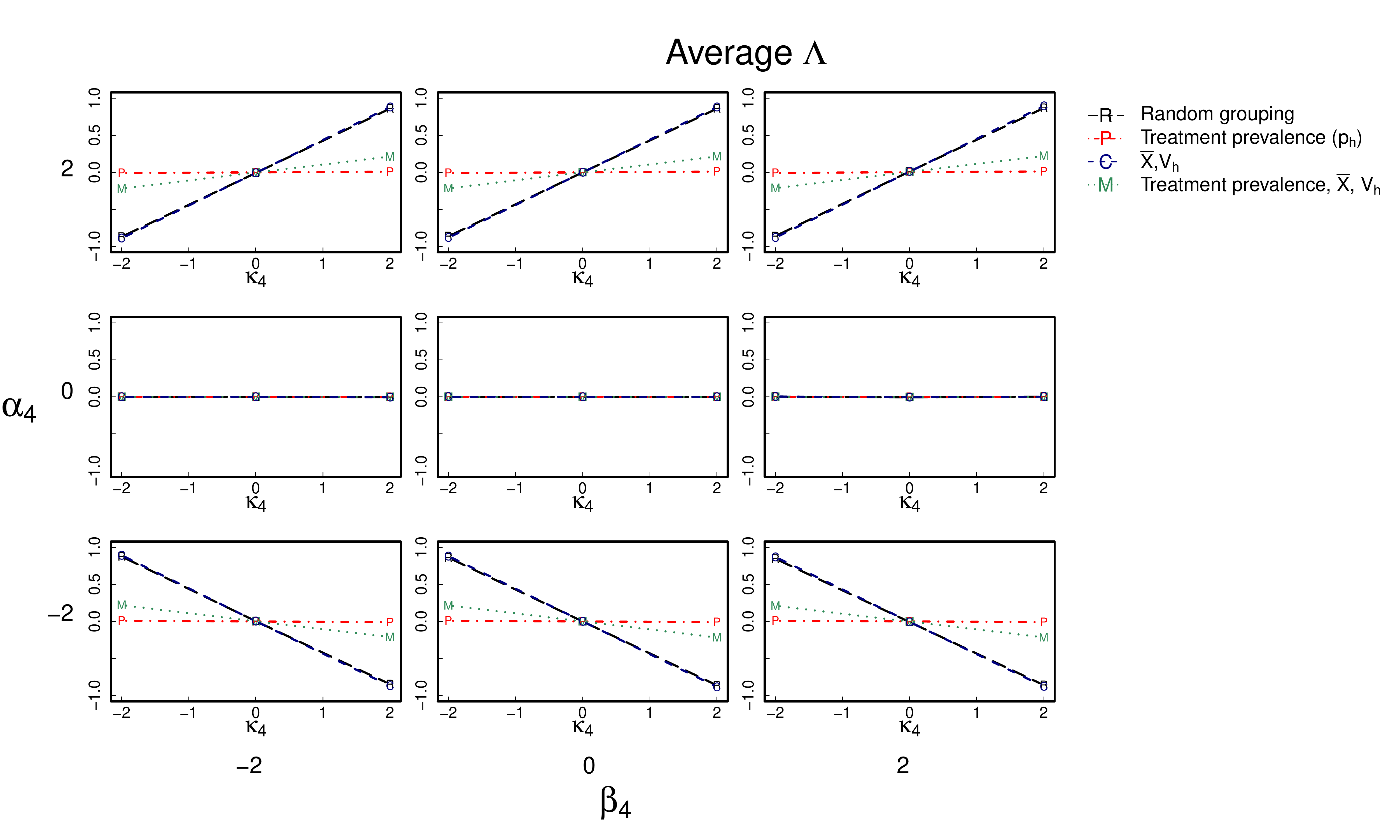} \\
\caption{\label{fig:sim121} Average bias in ATE (upper panel) and average estimated bias $\Lambda$ (lower panel) over $r=1000$ replicates when $(\beta^{\prime}, \kappa^{\prime}) = (2,1)$ \yl{when the propensity score models do not include random intercepts, i.e., $\hat{\tau}_{\text{(group, group)}}$ was used.}}	
\end{figure}

Figure~\ref{fig:sim112_norandom} shows increased bias across all grouping criteria compared to \yl{Figure~4} in the main text due to missing random intercepts. Figure~\ref{fig:sim121} shows bias in the ATE estimates when $(\beta^{\prime}, \kappa^{\prime}) = (2,1)$. Both figures show that $\Lambda$ captures bias pattern and the bias is minimized when clusters were partially grouped using treatment prevalence (red lines).

\begin{landscape}
\begin{table}[H]
\centering
\caption{\label{tab:alpha_2} Average absolute bias ($|$Bias$|$), standard error of the IPW estimator with propensity scores fixed (SE), and coverage rate of the estimates based on SE (Coverage) in simulations in \yl{Figure~5} under \yl{Equation (10)}.}
\resizebox{1.2\textwidth}{!}{\begin{tabular}{r|r||rrr|rrr|rrr|rrr|rrr|rrr}
\multicolumn{2}{c||}{(PS model, IPW)} & \multicolumn{3}{c|}{(full-RE, cluster)} & \multicolumn{3}{c|}{(full-RE, group)} & \multicolumn{3}{c|}{(full-RE, full)} & \multicolumn{3}{c|}{(group-RE, cluster)} & \multicolumn{3}{c|}{(group-RE, group)} & \multicolumn{3}{c}{(group-RE, full)} \\
  \hline
$\kappa_{4}$ & $\beta_{4}$ & $|$Bias$|$ & SE & Coverage & $|$Bias$|$ & SE & Coverage & $|$Bias$|$ & SE & Coverage & $|$Bias$|$ & SE & Coverage & $|$Bias$|$ & SE & Coverage & $|$Bias$|$ & SE & Coverage \\ 
  \hline
  \hline
  \multicolumn{2}{c}{$\alpha_{4} = -2$} \\
  \hline
\multirow{3}{*}{-2} & -2 &  0.087 & 0.094 & 0.938 & 0.127 & 0.142 & 0.971 & 0.577 & 0.134 & 0.001 & 0.059 & 0.097 & 0.992 & 0.074 & 0.146 & 1.000 & 0.079 & 0.426 & 0.999 \\ 
 & 0 &    0.085 & 0.094 & 0.936 & 0.101 & 0.132 & 0.985 & 0.057 & 0.112 & 0.997 & 0.057 & 0.097 & 0.988 & 0.064 & 0.136 & 0.998 & 0.061 & 0.281 & 1.000 \\ 
 & 2 & 0.086 & 0.094 & 0.942 & 0.082 & 0.155 & 1.000 & 0.537 & 0.157 & 0.000 & 0.058 & 0.097 & 0.993 & 0.060 & 0.160 & 1.000 & 0.066 & 0.412 & 0.999 \\ 
 \hline
\multirow{3}{*}{0} &  -2 & 0.045 & 0.063 & 0.980 & 0.087 & 0.123 & 0.990 & 0.535 & 0.127 & 0.000 & 0.044 & 0.066 & 0.979 & 0.056 & 0.127 & 0.999 & 0.061 & 0.374 & 1.000 \\ 
&  0 &  0.045 & 0.063 & 0.976 & 0.066 & 0.094 & 0.982 & 0.050 & 0.086 & 0.993 & 0.045 & 0.066 & 0.982 & 0.053 & 0.098 & 0.996 & 0.053 & 0.266 & 0.995 \\ 
&  2 &  0.044 & 0.063 & 0.978 & 0.060 & 0.109 & 0.999 & 0.576 & 0.126 & 0.000 & 0.044 & 0.066 & 0.987 & 0.057 & 0.112 & 0.999 & 0.065 & 0.376 & 1.000 \\ 
\hline
\multirow{3}{*}{2} &  -2 & 0.067 & 0.096 & 0.975 & 0.066 & 0.170 & 1.000 & 0.495 & 0.159 & 0.007 & 0.054 & 0.099 & 0.997 & 0.061 & 0.175 & 1.000 & 0.064 & 0.416 & 1.000 \\ 
&  0 &  0.067 & 0.096 & 0.968 & 0.063 & 0.137 & 1.000 & 0.077 & 0.114 & 0.980 & 0.058 & 0.099 & 0.996 & 0.064 & 0.142 & 1.000 & 0.061 & 0.287 & 1.000 \\ 
&  2 &  0.069 & 0.096 & 0.969 & 0.066 & 0.134 & 0.997 & 0.616 & 0.133 & 0.000 & 0.057 & 0.099 & 0.992 & 0.074 & 0.138 & 0.993 & 0.080 & 0.364 & 1.000 \\ 
   \hline
   \multicolumn{2}{c}{$\alpha_{4} = 0$} \\
   \hline
\multirow{3}{*}{-2} & -2 &  0.045 & 0.091 & 0.999 & 0.150 & 0.159 & 0.971 & 0.080 & 0.144 & 0.998 & 0.043 & 0.093 & 1.000 & 0.061 & 0.162 & 1.000 & 0.061 & 0.359 & 1.000 \\ 
 & 0 &  0.046 & 0.091 & 0.998 & 0.145 & 0.122 & 0.910 & 0.057 & 0.107 & 0.999 & 0.045 & 0.093 & 0.999 & 0.052 & 0.124 & 0.999 & 0.052 & 0.257 & 1.000 \\ 
 & 2 & 0.045 & 0.091 & 0.998 & 0.145 & 0.159 & 0.968 & 0.080 & 0.144 & 0.997 & 0.045 & 0.093 & 0.999 & 0.061 & 0.162 & 0.999 & 0.061 & 0.379 & 1.000 \\ 
 \hline
\multirow{3}{*}{0} &  -2 &  0.042 & 0.059 & 0.976 & 0.146 & 0.131 & 0.930 & 0.074 & 0.127 & 0.995 & 0.040 & 0.061 & 0.981 & 0.055 & 0.134 & 1.000 & 0.056 & 0.358 & 1.000 \\ 
&  0 &  0.042 & 0.059 & 0.973 & 0.147 & 0.084 & 0.620 & 0.046 & 0.081 & 0.996 & 0.041 & 0.061 & 0.983 & 0.045 & 0.085 & 0.998 & 0.046 & 0.252 & 0.997 \\ 
&  2 & 0.044 & 0.059 & 0.975 & 0.145 & 0.131 & 0.918 & 0.074 & 0.126 & 0.989 & 0.042 & 0.061 & 0.978 & 0.057 & 0.134 & 1.000 & 0.057 & 0.391 & 1.000 \\ 
\hline
\multirow{3}{*}{2} &  -2 &0.047 & 0.096 & 0.997 & 0.152 & 0.162 & 0.970 & 0.080 & 0.145 & 0.993 & 0.046 & 0.098 & 0.997 & 0.062 & 0.165 & 1.000 & 0.062 & 0.380 & 1.000 \\ 
 & 0 & 0.048 & 0.096 & 0.999 & 0.148 & 0.126 & 0.916 & 0.059 & 0.107 & 0.996 & 0.046 & 0.097 & 0.999 & 0.054 & 0.128 & 1.000 & 0.053 & 0.280 & 1.000 \\ 
 & 2 &   0.044 & 0.096 & 0.999 & 0.142 & 0.162 & 0.975 & 0.083 & 0.144 & 0.993 & 0.044 & 0.097 & 0.999 & 0.061 & 0.165 & 1.000 & 0.061 & 0.399 & 1.000 \\ 
 \hline
  \multicolumn{2}{c}{$\alpha_{4} = 2$} \\
  \hline
\multirow{3}{*}{-2} & -2 & 0.084 & 0.094 & 0.924 & 0.080 & 0.155 & 1.000 & 0.538 & 0.157 & 0.001 & 0.058 & 0.097 & 0.984 & 0.059 & 0.161 & 1.000 & 0.067 & 0.413 & 1.000 \\ 
 & 0 &  0.084 & 0.094 & 0.944 & 0.098 & 0.132 & 0.979 & 0.057 & 0.112 & 0.996 & 0.056 & 0.097 & 0.988 & 0.062 & 0.136 & 0.997 & 0.061 & 0.293 & 1.000 \\ 
&  2 &  0.084 & 0.094 & 0.939 & 0.121 & 0.142 & 0.960 & 0.569 & 0.134 & 0.000 & 0.059 & 0.097 & 0.980 & 0.077 & 0.147 & 0.995 & 0.082 & 0.406 & 0.999 \\ 
\hline
\multirow{3}{*}{0} & -2 & 0.044 & 0.063 & 0.977 & 0.058 & 0.109 & 0.997 & 0.573 & 0.126 & 0.000 & 0.044 & 0.066 & 0.985 & 0.055 & 0.112 & 0.998 & 0.063 & 0.388 & 1.000 \\ 
  & 0 &   0.045 & 0.063 & 0.975 & 0.067 & 0.094 & 0.980 & 0.048 & 0.086 & 0.994 & 0.045 & 0.066 & 0.985 & 0.051 & 0.098 & 0.996 & 0.051 & 0.251 & 0.997 \\ 
&  2 & 0.045 & 0.063 & 0.979 & 0.084 & 0.123 & 0.991 & 0.529 & 0.127 & 0.000 & 0.044 & 0.066 & 0.983 & 0.056 & 0.127 & 0.999 & 0.062 & 0.387 & 1.000 \\ 
\hline
\multirow{3}{*}{2} &  -2 & 0.070 & 0.096 & 0.976 & 0.069 & 0.134 & 0.998 & 0.610 & 0.133 & 0.000 & 0.059 & 0.099 & 0.989 & 0.077 & 0.139 & 0.993 & 0.085 & 0.403 & 0.999 \\ 
&  0 & 0.069 & 0.096 & 0.976 & 0.059 & 0.137 & 1.000 & 0.075 & 0.113 & 0.989 & 0.056 & 0.099 & 0.990 & 0.064 & 0.142 & 0.997 & 0.062 & 0.293 & 1.000 \\ 
&  2 & 0.071 & 0.096 & 0.970 & 0.066 & 0.170 & 1.000 & 0.495 & 0.159 & 0.004 & 0.058 & 0.099 & 0.987 & 0.060 & 0.176 & 1.000 & 0.062 & 0.414 & 1.000 \\  
 \hline
\end{tabular}}
\end{table}
\end{landscape}

\subsection{Comparison with existing approaches}
\label{ssec:compare}

\begin{figure}%
	\centering
	\includegraphics[width=\textwidth]{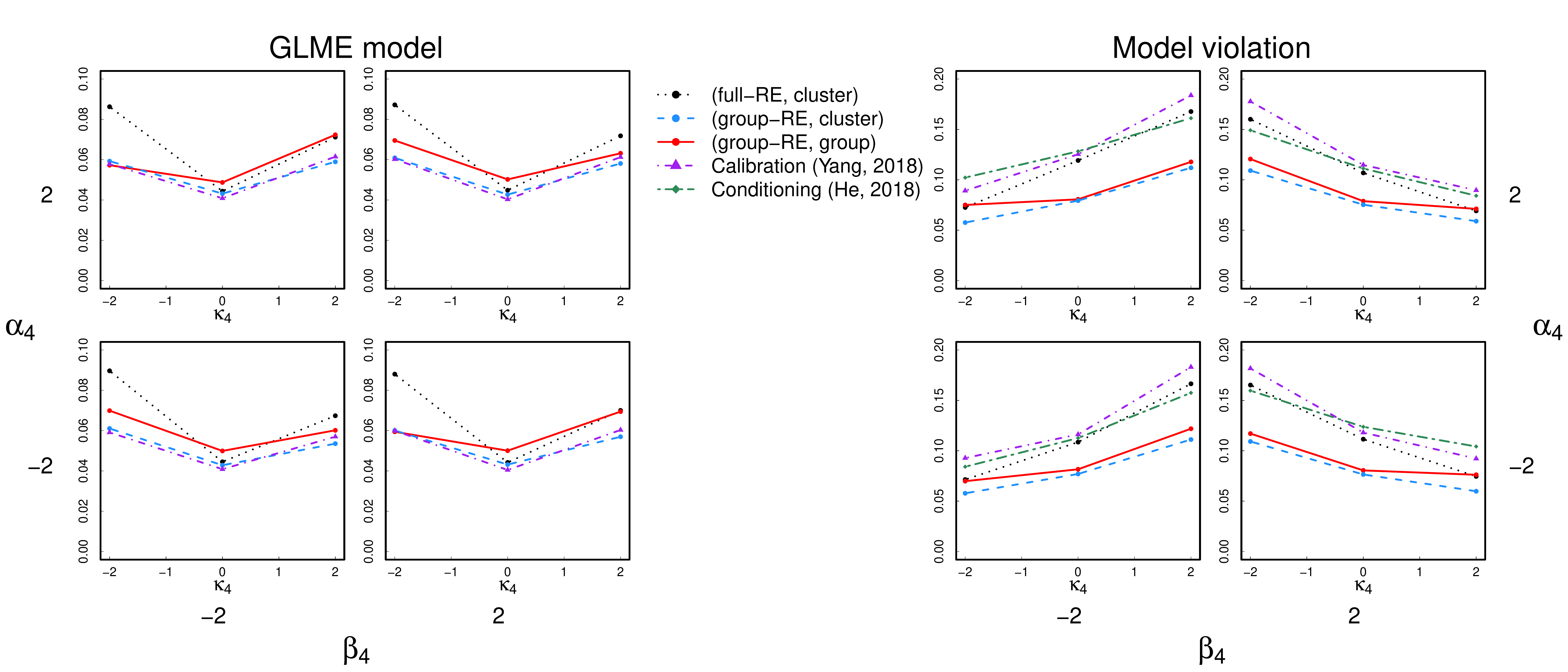}
\caption{\label{fig:cali} 
The left panel shows methods' performance under correct GLME model specification for the calibration~\citep{yang2018propensity} and conditioning~\citep{he2018inverse} approaches. \yl{On the other hand, the right panel shows the performance under the different data generating model that violates the model specification for the two existing methods}.}
\end{figure}

We compared the aforementioned three IPW estimators to the two existing methods of calibration~\citep{yang2018propensity} and conditioning by sufficient statistics~\citep{he2018inverse} in Figure~\ref{fig:cali}. The left panel illustrates the same results presented in \yl{Figure 5} in the main text only under the presence of confounding ($\alpha_{4}, \beta_{4} \neq 0$), and we add the results of the \citeauthor{yang2018propensity}'s calibration method (purple lines) on it. We omitted the results of the conditioning method in this panel because they exhibit much larger absolute bias than the other five (although still smaller bias than those from (full, full) estimators). When Equation (10) in the main text is correct, the \citeauthor{yang2018propensity}'s calibration and (group-RE, cluster) estimators perform the best, with essentially equivalent performance between the two. 

The propensity score model and outcome model generated by Equation (10) in the main text conforms to the generalized linear mixed models that the calibration and conditioning methods assume. To explore robustness of the methods to violation of this assumption, we also considered a simple scenario where a ``random effect" from $U_{h}$ interacts with the observed covariates of $X_{hk}$ in the outcome model, thus violating the parametric model assumption underlying the approaches. This is a plausible scenario in that the influence of individual characteristics (e.g., math scores from the previous exam) could be easily modified by contextual characteristics (e.g., different evaluation criteria across states). This type of model is formalized in  Equation~\eqref{eq:interaction}:
\begin{eqnarray}
\label{eq:interaction}
Y_{hk} & =&   \beta_{0,h} +  \beta_{1} \bar{X}_{h}  +  \left\{ \beta_{2} (X_{hk} - \bar{X}_{h}) \times \beta_{4}  (U_{h} - \bar{U})  \right\} +  \beta_{3} V_{h}  + \\ & &Z_{hk} \left[ \kappa_{0,h} + \kappa_{1} \bar{X}_{h}  +  \left\{ \kappa_{2} (X_{hk} - \bar{X}_{h}) \times \kappa_{4}  (U_{h} - \bar{U})  \right\} + \kappa_{3} V_{h}    \right]  + \epsilon_{hk}. \nonumber
\end{eqnarray}
The right panel of Figure~\ref{fig:cali} shows that when compared to the calibration and conditioning approaches, the (\yl{group-RE}, group) and (\yl{group-RE}, cluster) estimators have smaller absolute bias when the interaction terms influence the overall outcome distribution (i.e., non-zero $\beta_{4}$) or the treatment effects (i.e., non-zero $\kappa_{4}$). This implies that the partially pooled propensity scores are less sensitive to this key assumption that underlies the existing approaches.

\subsection{\yl{Comparison to the latent class approach}}

In this section, we compared our partial pooling using the treatment prevalence to latent model approach in~\cite{kim2015mixture, kim2016causal}. The latent model approach detects latent classes among clusters in the treatment assignment model and/or the outcome model. In our contexts, we only fitted a latent class model for the treatment assignment to partially pool group of clusters with similarity. 
Our main aim here is to compare the partial pooling by the latent class model and that by the treatment prevalence.
After pooling the group of clusters using either method, we estimated the propensity scores within each group of clusters and evaluated the average causal effect using the group-weighted IPW estimator. We generated the following treatment assignment and potential outcome model as introduced in Section~\ref{sec:simresult} with $(\beta^{\prime}, \kappa^{\prime}) = (1,2)$. 

\begin{figure}[H]
\centering
\includegraphics[width=0.8\textwidth]{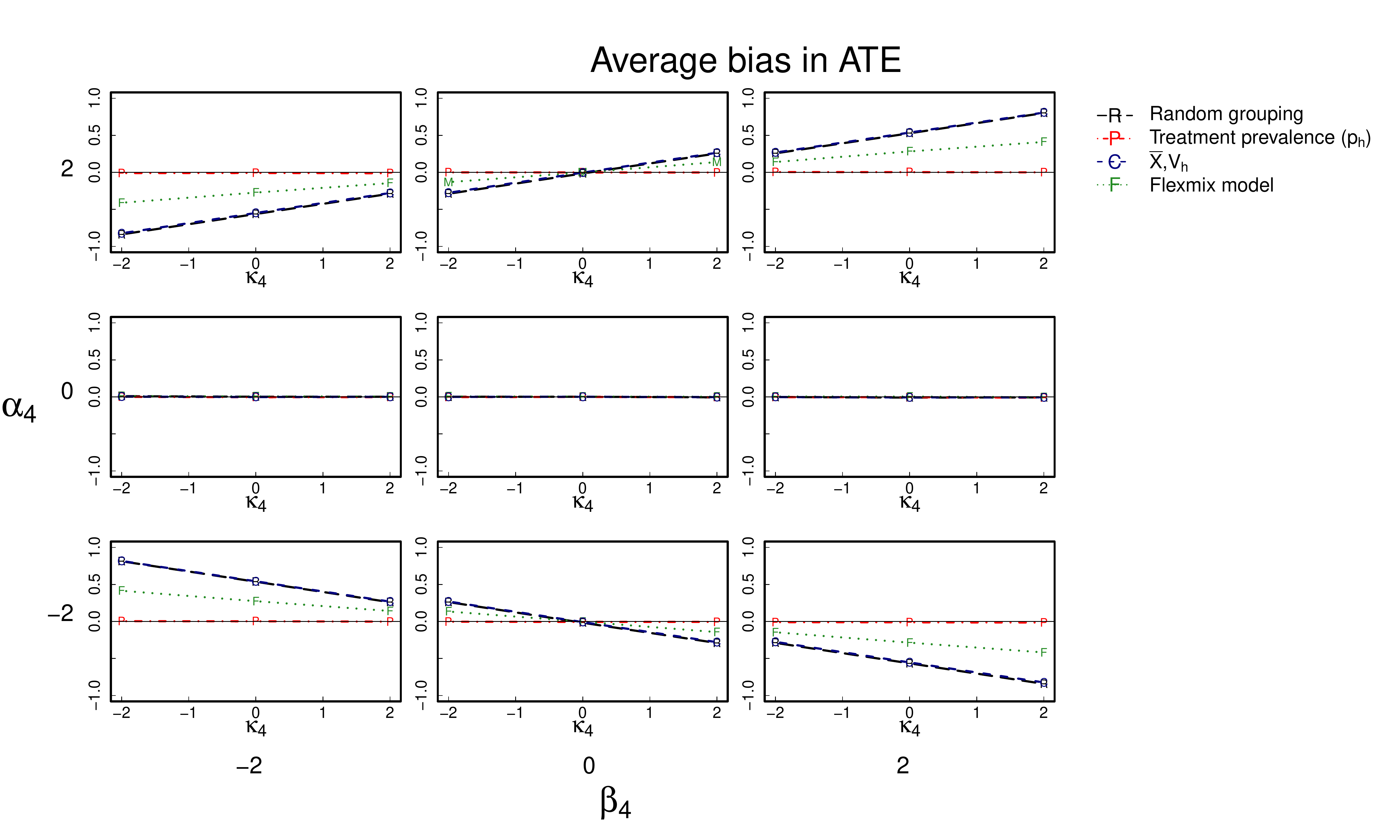}
\caption{\label{fig:flexmix}  Average bias in ATE over $r=1000$ replicates when different (\yl{group-RE}, group) estimators are used. For partially pooled propensity score models, $H=200$ clusters are selectively pooled into up-to $G=10$ groups (R) Randomly, (P) by minimizing the within-group distance of the treatment prevalence within group, (C) by minimizing the within-group distance of the observed covariates, or (F) using  a flexible mixture model (\texttt{Flexmix} in \texttt{R}) to pull latent classes of clusters with respect to propensity scores. Across over all the scenarios, the (R) and (C) curves sit right on top of one another. In case of (F), the maximum number of ``latent classes" is $G=10$, but the actual number of classes can be less than ten.}
\end{figure}

Figure~\ref{fig:flexmix} presents the average bias in ATE using four different cluster pooling methods. Since we used the same propensity score model upon the pooled cluster groups and the same IPW estimator, the difference in the methods can be attributed only to the effectiveness of each partial pooling method. The latent class approach (Flexmix) improves the estimation, but not as much as the partial pooling using the treatment prevalence (red lines). This might be because when the latent features among clusters are continuous (e.g., $U_{h} \sim \mathcal{U}(-2, 2)$), the latent class models may fail to capture the discrete number of clusters. In fact, even though we specified the maximum number of latent classes as $G=10$, less than ten classes were often generated.

\section{Additional results from ECLS-K example}
\label{sec:ecls}

\begin{figure}[H]
\centering
\includegraphics[width=0.7\textwidth]{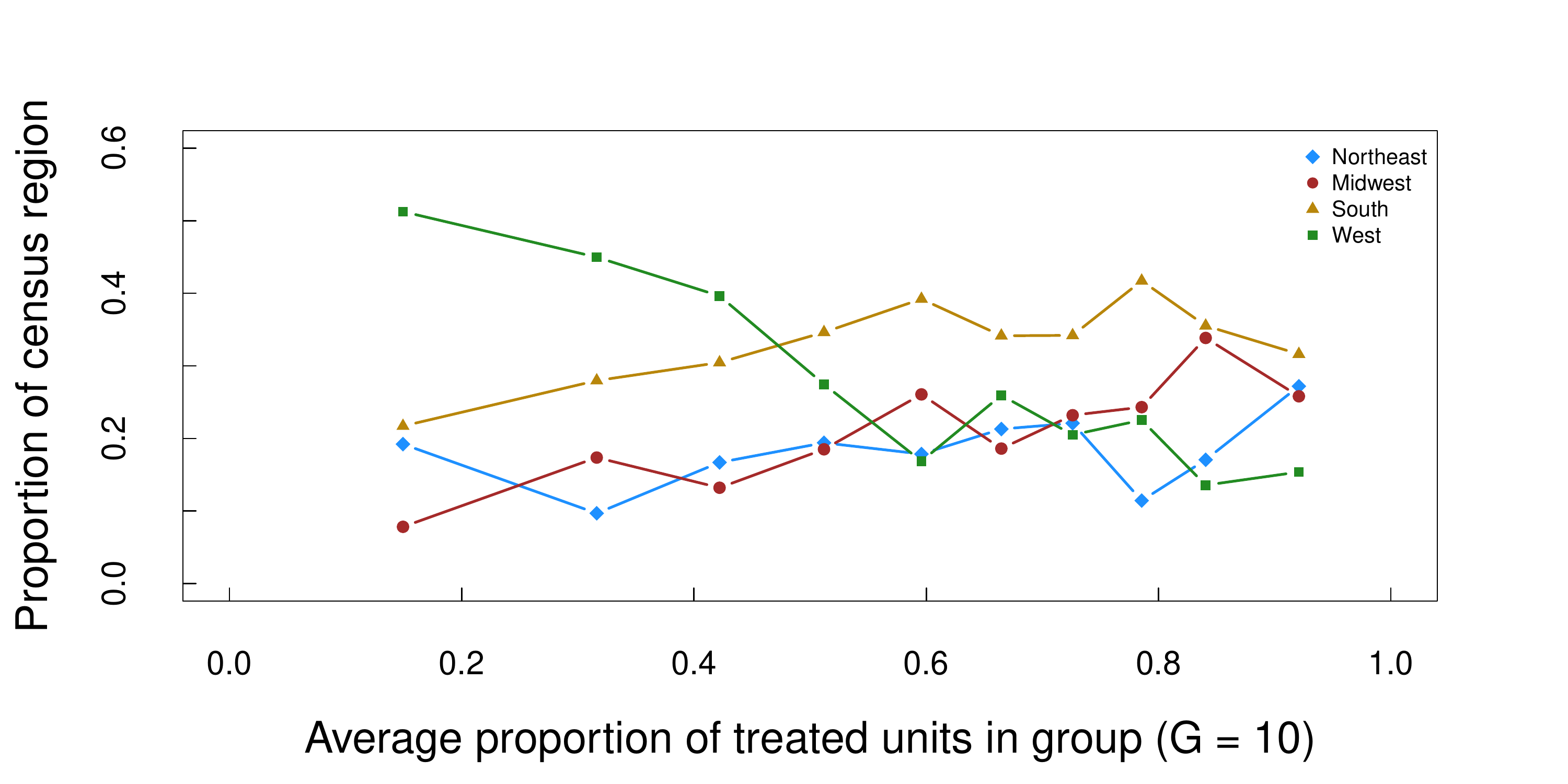}
\caption{\label{fig:region}  Proportion of four different census regions for ten different cluster groups identified by PAM using \texttt{ClusterR} package. As the average treatment prevalence in the group increases the proportion of schools from West noticeably decreases.}
\end{figure}

Figure~\ref{fig:region} illustrates the distributions of the cluster-level characteristic of census region according to the average treatment prevalence in the cluster group identified by the PAM algorithm.

\begin{table}[H]
\centering
\caption{IPW estimates (Estimate), standard errors with propensity scores fixed (SE), and confidence intervals constructed through $r=1000$ bootstrap samples (95\% CI) for each weighting scheme (unweighted/(\yl{full-RE}, full)/(\yl{group-RE}, group)) and with/without regional information in a propensity score model.}
\resizebox{0.9\textwidth}{!}{\begin{tabular}{rr|rrr}
  \hline
  & & Estimate & SE & 95\% CI  \\ 
  \hline
 \multicolumn{2}{r|}{Unweighted effect} & 4.36 & 0.29 & [3.91, 4.85] \\ 
\hline
\multirow{3}{*}{With regional information in PS model}& (full-RE, full) & 2.30 & 0.22 & [1.93, 2.71] \\ 
&  (group-RE, group) & 1.62 & 0.29 & [1.14, 2.12] \\ 
&  (group-RE, cluster) & 1.84 & 0.26 & [1.38, 2.36] \\ 
  \hline
\multirow{3}{*}{Without regional information in PS model}&  (full-RE, full) & 2.36 & 0.22 & [2.01, 2.79] \\ 
& (group-RE, group) & 1.62 & 0.29 & [1.16, 2.12] \\ 
& (group-RE, cluster) & 1.86 & 0.26 & [1.40, 2.37] \\ 
   \hline
\end{tabular}}
\end{table}

\yl{In the main analysis, we considered the following demographic, individual-level covariates: gender, age, height, weight, race/ethnicity (indicators of a black and Hispanic child), an indicator of having two parents, family types (indicators of two-parent only, one parent and siblings, and one parent only), and an indicator of a biological mother. Moreover, we included an indicator of using English at home and motor skills as education-related individual-level covariates. One caveat here is that these two education-related covariates, as well as the child's height and weight, were measured at the base-year of Kindergarten. Also, we assumed that the treatment (i.e., types of pre-school education) does not affect these variables measured at the beginning of Kindergarten but may be correlated with each of them only through each variable's pre-treatment status, e.g., weight at age 4. More details on the conditioning covariates can be found in~\cite{dong2020using}.}

\yl{As cluster-level (i.e., school-level) covariates, census region (Northeast, Midwest, South, and West) and location (Central city, Urban fringe/large town, and small-town/rural) are included. Often geographic information is known in many survey data, but for comparison purpose, we hypothetically assumed that we had not observed these two cluster-level covariates. Table~\ref{tab:without_region} illustrate the covariate balance table for observed and (presumably) unobserved covariates with and without weighting.}

\begin{table}[H]
\centering
\caption{\label{tab:without_region} Covariate balance table when observations are not weighted by propensity scores (unweighted PS), weighted by partially pooled propensity score (PS), and weighted by fully pooled propensity score (PS).  Both partially pooled propensity score model and fully pooled propensity score model omitted cluster-level variable of census region and location (Model 2)*, \yl{both with random effects included in the model}.}
\resizebox{\textwidth}{!}{\begin{tabular}{r||lll|lll|lll}
  \hline
  & \multicolumn{3}{c|}{Unweighted PS} 
  & \multicolumn{3}{c|}{Weighted by partially pooled PS} & 
  \multicolumn{3}{c}{Weighted by fully pooled PS} \\
Variable & Parental care & Center-based & SMD & Parental care & Center-based & SMD & Parental care & Center-based & SMD \\ 
  \hline

  Child's math score  &     24.28 (8.32) & 28.64 (9.61) &  0.485 &   26.12 (9.09) &   27.74 (9.42) &  0.176 &   25.65 (8.75) &   28.01 (9.54) &  0.258 \\ 
  Female  &     0.47 (0.50) &  0.49 (0.50) &  0.039 &    0.48 (0.50) &    0.49 (0.50) &  0.005 &    0.48 (0.50) &    0.49 (0.50) &  0.010 \\ 
  Black  &     0.11 (0.31) &  0.12 (0.33) &  0.040 &    0.12 (0.32) &    0.12 (0.32) &  0.008 &    0.12 (0.32) &    0.12 (0.32) &  0.006 \\ 
  Hispanic  &     0.26 (0.44) &  0.13 (0.33) &  0.336 &    0.17 (0.37) &    0.17 (0.38) &  0.008 &    0.17 (0.38) &    0.17 (0.37) &  0.014 \\ 
  Two parents  &    0.83 (0.38) &  0.82 (0.38) &  0.013 &    0.83 (0.37) &    0.82 (0.38) &  0.028 &    0.83 (0.38) &    0.83 (0.38) &  0.010 \\ 
  Family type  &    &   &  0.211 &   &   &  0.032 &   &   &  0.013 \\ 
  2 parents + siblings &  2053 (76.0)  &  4161 (71.0)  &  &  6282.6 (73.6)  &  6198.7 (72.4)  &  &  5754.8 (73.1)  &  6029.7 (72.6)  &  \\ 
  2 parents only &   183 ( 6.8)  &   661 (11.3)  &  &   830.3 ( 9.7)  &   844.4 ( 9.9)  &  &   775.8 ( 9.9)  &   827.9 (10.0)  &  \\ 
  1 parent + siblings  &   326 (12.1)  &   629 (10.7)  &  &   908.6 (10.6)  &   984.7 (11.5)  &  &   856.7 (10.9)  &   920.1 (11.1)  &  \\ 
  1 parent only  &   93 ( 3.4)  &   348 ( 5.9)  &  &   424.5 ( 5.0)  &   434.5 ( 5.1)  &  &   394.0 ( 5.0)  &   432.8 ( 5.2)  &  \\ 
  Other &      45 ( 1.7)  &    58 ( 1.0)  &  &    90.9 ( 1.1)  &    99.9 ( 1.2)  &  &    94.3 ( 1.2)  &    97.3 ( 1.2)  &  \\ 
  Biological mother  &   0.96 (0.20) &  0.95 (0.22) &  0.047 &    0.95 (0.22) &    0.95 (0.21) &  0.009 &    0.95 (0.21) &    0.95 (0.21) &  0.010 \\ 
 English use at home  &   0.80 (0.40) &  0.91 (0.28) &  0.322 &    0.88 (0.33) &    0.88 (0.33) &  0.003 &    0.87 (0.33) &    0.88 (0.33) &  0.019 \\ 
  Motor skills  &   11.91 (3.14) & 12.72 (2.86) &  0.270 &   12.49 (2.96) &   12.48 (2.97) &  0.003 &   12.45 (2.97) &   12.48 (3.00) &  0.009 \\ 
  Height  &   44.56 (2.17) & 44.78 (2.14) &  0.102 &   44.69 (2.14) &   44.70 (2.16) &  0.005 &   44.69 (2.16) &   44.71 (2.15) &  0.011 \\ 
  Weight &   46.01 (8.89) & 46.40 (8.34) &  0.045 &   46.21 (8.68) &   46.28 (8.40) &  0.008 &   46.32 (8.85) &   46.30 (8.33) &  0.002 \\ 
  Age  &     68.53 (4.42) & 68.62 (4.21) &  0.021 &   68.52 (4.29) &   68.60 (4.22) &  0.018 &   68.50 (4.33) &   68.60 (4.22) &  0.023 \\ 
  \textbf{Census region*}  &    &   &  0.211 &   &   &  0.026 &   &   &  0.080 \\ 
Northeast &    473 (17.5)  &  1165 (19.9)  &  &  1696.1 (19.9)  &  1620.2 (18.9)  &  &  1503.5 (19.1)  &  1613.3 (19.4)  &  \\ 
 Midwest &    549 (20.3)  &  1465 (25.0)  &  &  2025.6 (23.7)  &  2024.0 (23.6)  &  &  1761.0 (22.4)  &  2019.2 (24.3)  &  \\
  South &    894 (33.1)  &  2034 (34.7)  &  &  2852.1 (33.4)  &  2894.1 (33.8)  &  &  2632.4 (33.4)  &  2855.5 (34.4)  &  \\ 
   West &   784 (29.0)  &  1193 (20.4)  &  &  1963.0 (23.0)  &  2023.9 (23.6)  &  &  1978.6 (25.1)  &  1819.8 (21.9)  &  \\ 
 \textbf{Location*}   &   &   &  0.190 &   &   &  0.041 &   &   &  0.124 \\ 
  Central city &   1110 (41.1)  &  2363 (40.3)  &  &  3401.0 (39.8)  &  3509.6 (41.0)  &  &  3118.6 (39.6)  &  3402.4 (41.0)  &  \\ 
Large town &    1000 (37.0)  &  2594 (44.3)  &  &  3552.7 (41.6)  &  3594.1 (42.0)  &  &  3137.3 (39.8)  &  3586.8 (43.2)  &  \\ 
  Small town &       590 (21.9)  &   900 (15.4)  &  &  1583.1 (18.5)  &  1458.5 (17.0)  &  &  1619.6 (20.6)  &  1318.5 (15.9)  &  \\ 
  Sum of weights & 2700 &  5857 &  & 8536.83 & 8562.19 &  & 7875.49 & 8307.70 &  \\ 
   \hline
\end{tabular}}
\end{table}

\end{document}